\documentstyle[12pt,psfig]{article}

\begin{document}

\newcommand{\ep}{equivalence principle~}
\newcommand{\Ch}{Chandrasekhar~}
\newcommand{\Chp}{Chandrasekhar}
\newcommand{\Sc}{Schwarzschild~}
\newcommand{\Scp}{Schwarzschild}
\newcommand{\Sw}{Schwarzschild~}
\newcommand{\Swp}{Schwarzschild}
\newcommand{\Sch}{Schr{\"{o}}dinger~}
\newcommand{\Schp}{Schr{\"{o}}dinger}
\newcommand{\OVp}{Oppenheimer--Volkoff}
\newcommand{\OV}{Oppenheimer--Volkoff~}
\newcommand{\GR}{General Relativity~}
\newcommand{\GT}{General Theory of Relativity~}
\newcommand{\GRp}{General Relativity}
\newcommand{\GTp}{General Theory of Relativity}
\newcommand{\STp}{Special Theory of Relativity}
\newcommand{\ST}{Special Theory of Relativity~}
\newcommand{\Lt}{Lorentz transformation~}
\newcommand{\Ltp}{Lorentz transformation}
\newcommand{\rel}{relativistic~}
\newcommand{\relp}{relativistic}
\newcommand{\msun}{M_{\odot}}
\newcommand{\eos}{equation of state~}
\newcommand{\eoss}{equations of state~}
\newcommand{\eossp}{equations of state}
\newcommand{\eosp}{equation of state}
\newcommand{\Eos}{Equation of state}
\newcommand{\Eosp}{Equation of state}
\newcommand{\beqn}{\begin{eqnarray}}
\newcommand{\eeqn}{\end{eqnarray}}
\newcommand{\nonum}{\nonumber \\}
\newcommand{\walecka}{$\sigma,\omega,\rho$~}
\newcommand{\waleckap}{$\sigma,\omega,\rho$}
\newcommand{\bbar}[1] {\mbox{$\overline{#1}$}} 
\newcommand{\courtesy}{~Reprinted with permission of Springer--Verlag 
New York; copyright 1997}
\newcommand{\oo}{{\"{o}}}
\newcommand{\au}{{\"{a}}}


\newcommand{\approxlt} {\mbox {$\stackrel{{\textstyle<}} {_\sim}$}}
\newcommand{\approxgt} {\mbox {$\stackrel{{\textstyle>}} {_\sim}$}}
\newcommand{\mearth} {\mbox {$M_\oplus$}}
\newcommand{\rearth} {\mbox {$R_\oplus$}}
\newcommand{\eo} {\mbox{$\epsilon_0$}}
\newcommand{\bfalpha} {\mbox{\mbox{\boldmath$\alpha$}}}
\newcommand{\bfgamma} {\mbox{\mbox{\boldmath$\gamma$}}}
\newcommand{\bfrho} {\mbox{\mbox{\boldmath$\rho$}}}
\newcommand{\bfsigma} {\mbox{\mbox{\boldmath$\sigma$}}}
\newcommand{\bftau} {\mbox{\mbox{\boldmath$\tau$}}}
\newcommand{\bfLambda} {\mbox{\mbox{\boldmath$\Lambda$}}}
\newcommand{\bfpi} {\mbox{\mbox{\boldmath$\pi$}}}
\newcommand{\bfomega} {\mbox{\mbox{\boldmath$\omega$}}}
\newcommand{\bp} {\mbox{\mbox{\boldmath$p$}}}
\newcommand{\br} {\mbox{\mbox{\boldmath$r$}}}
\newcommand{\bx} {\mbox{\mbox{\boldmath$x$}}}
\newcommand{\bv} {\mbox{\mbox{\boldmath$v$}}}
\newcommand{\bu} {\mbox{\mbox{\boldmath$u$}}}
\newcommand{\bk} {\mbox{\mbox{\boldmath$k$}}}
\newcommand{\bA} {\mbox{\mbox{\boldmath$A$}}}
\newcommand{\bB} {\mbox{\mbox{\boldmath$B$}}}
\newcommand{\bF} {\mbox{\mbox{\boldmath$F$}}}
\newcommand{\bI} {\mbox{\mbox{\boldmath$I$}}}
\newcommand{\bJ} {\mbox{\mbox{\boldmath$J$}}}
\newcommand{\bK} {\mbox{\mbox{\boldmath$K$}}}
\newcommand{\bP} {\mbox{\mbox{\boldmath$P$}}}
\newcommand{\bS} {\mbox{\mbox{\boldmath$S$}}}
\newcommand{\bdel} {\mbox{\mbox{\boldmath$\bigtriangledown$}}}

\newcommand{\eps} {\mbox {$\epsilon$}}
\newcommand{\gpercm} {\mbox {${\rm g}/{\rm cm}^{3}$}}
\newcommand{\rhon} {\mbox {$\rho_{0}$}}
\newcommand{\rhoc} {\mbox {$\rho_{c}$}}
\newcommand{\fmm} {\mbox {${\rm fm}^{-3}$}}
\newcommand{\bag} {\mbox {$B^{1/4}$}}

\newcommand{\fraca} {\mbox {$ \frac{1}{2}       $}}
\newcommand{\fracb} {\mbox {$ \frac{3}{2}       $}}
\newcommand{\fracc} {\mbox {$ \frac{1}{4}       $}}
\newcommand{\fraccc} {\mbox {$ \frac{1}{3}       $}}

\newcommand{\tit}
{First Order Kaon Condensate}

\newcommand{\auth} {Norman K. Glendenning and J\"urgen Schaffner-Bielich}
\newcommand{\lbl}{LBNL-42330}
\newcommand{\dateofdoc}{October 8, 1997}
\newcommand{\adr} 
{Nuclear Science Division \& 
Institute for Nuclear and Particle Astrophysics,
  Lawrence Berkeley  National Laboratory,
   MS: 70A-3307 \\ Berkeley, California 94720}

\newcommand{\doe}
{This work was supported by the
Director, Office of Energy Research,
Office of High Energy
and Nuclear Physics,
Division of Nuclear Physics,
of the U.S. Department of Energy under Contract
DE-AC03-76SF00098.}

\newcommand{\ect}{A part of this work was done at the ECT*,
Villa Tambosi, Trento, Italy.}


\begin{titlepage}
\begin{center}
\parbox{3in}{\begin{flushleft}Preprint \end{flushleft}}%
\parbox{3in}{\begin{flushright} \lbl  \end{flushright}}
~\\[7ex]

\renewcommand{\thefootnote}{\fnsymbol{footnote}}
\setcounter{footnote}{1}

\begin{Large}
\tit {\footnote{\doe}}\\[2ex]
\end{Large}
\renewcommand{\thefootnote}{\fnsymbol{footnote}}
\setcounter{footnote}{1}
~~\footnotetext{\tiny{[nkg/papers/kaon.tex,  \today} }

\begin{large}
\auth\\[3ex]
\end{large}
\adr\\[3ex]
\dateofdoc \\[3ex]
\end{center}
\begin{figure}[htb]
\vspace{.8in}
\begin{center}
\leavevmode
\hspace{-.2in}
\psfig{figure=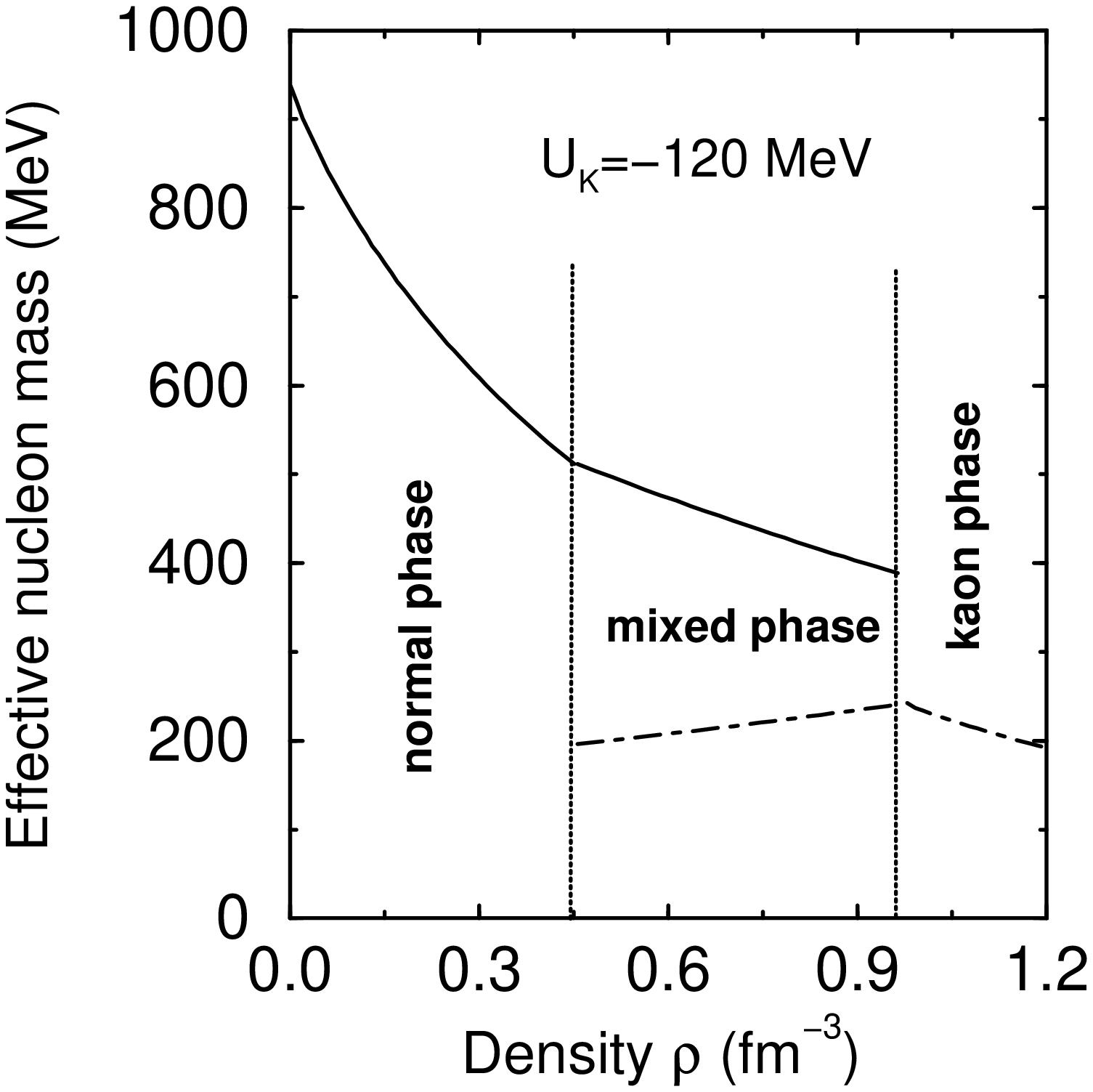,height=3.2in}
\end{center}
\end{figure}


\end{titlepage}

\clearpage


\begin{center}
\begin{Large}
\tit \\[7ex]
\end{Large}

\begin{large}
\auth \\[2ex]
\end{large}
\adr
\end{center}


\begin{abstract}
First order Bose condensation in asymmetric nuclear matter and in
neutron stars is studied, with particular reference to kaon condensation.
 We demonstrate
 explicitly why the Maxwell construction fails to
 assure equilibrium in multicomponent substances.
  Gibbs conditions and conservation laws 
require that  for phase equilibrium, 
  the charge density must have opposite sign in the
  two phases of isospin asymmetric
  nuclear matter. 
  The mixed phase will therefore form a Coulomb
  lattice with the rare phase occupying lattice sites in the dominant
  phase.  Moreover, the kaon condensed phase differs from the normal phase,
    not by the mere presence of kaons in the first, 
      but  also by a difference in the
        nucleon effective masses. The mixed phase region, which occupies
        a large radial extent amounting to some kilometers in our model
        neutron stars, is thus highly heterogeneous.
                It  should
                be particularly interesting in connection with the
                pulsar glitch phenomenon as well as transport properties.

\end{abstract}

\setcounter{footnote}{0}


\section{Introduction}

Many phase transitions may occur in superdense matter.  Among the possible new
phases that have been considered over the past few years are pion and kaon
condensed and quark deconfined matter.  Transitions from the normal to any of
these high-density phases may be of first or second order.  The order depends
in part on the strength of coupling constants.  If of first order, especially
interesting phenomena occur in isospin asymmetric nuclear matter, including
spatially ordered regions of the normal and new phase in the range of densities
for which both phases are in equilibrium \cite{glen91:d,glen91:a}.

In early work on pion condensation, the region of phase coexistence was found
by use of the Maxwell construction (sometimes with reference to Van der Waal's
equation of state; 
cf. Refs.\ \cite{hartle75:a,weise75:a,migdal79:a,migdal79:b,weise79:a}).
The Maxwell construction is valid for simple substances---those 
with {\sl only 
one} independent component, like water, or pure neutron matter (which nowhere
exists).  However, as used\footnote{To be sure, the Maxwell construction could
  be generalized. But ``equal areas'' would be replaced by ``equal volumes'',
  and ``tangent slope'' by ``tangent surface'' in a space of $n+1$ dimensions,
  where $n$ denotes the number of independent components.}, 
this construction can assure that only one chemical potential is common to the
two phases, whereas asymmetric nuclear matter, such as neutron star matter, has
two independent components (the baryon and electric charge).  Consequently the
construction cannot satisfy Gibbs criteria that all chemical potentials as well
as pressure and temperature be common to both phases in equilibrium.  In short,
the states studied were unstable.

Similarly the deconfinement transition was treated in early work beginning in
the 1970's with the same assumptions and methods and without regard to
equilibrium (in some cases without regard to beta equilibrium in the pure
phases and in others without regard to phase equilibrium).  
The deconfinement
phase transition for $\beta$ stable matter has been treated recently in some
detail taking account of equilibrium in all phases
  \cite{glen91:d,glen91:a,heiselberg93:a,pandharipande94:a,%
  glen95:c,glen95:e,glen97:c}. 
(The Gibbs criteria was used in Ref. 
\cite{heinz87:a,Greiner87} in heavy-ion physics where it was essential
for the distillation of strangeness.)

The possibility of kaon condensation was discussed already some years ago in
the context of hyperonized neutron star matter \cite{glen85:b}.  But the real
impetus for the recent interest was provoked by the paper of Kaplan and Nelson
\cite{KN86} who suggested the interaction of the K$^-$ with the nuclear
medium may reduce its mass sufficiently, so that, as a boson, it may replace
electrons as the neutralizing agent in charge neutral matter.  
The Maxwell construction was again used in previous work on first
order kaon
condensation \cite{Brown92,Thorsson94,Fujii96,Li97}.

In the present
paper we study the kaon condensed phase as a {\sl first order} phase transition
in neutron star matter.  This will serve as a general example of a Bose
condensate (whether pion or kaon).  Unlike the previously cited work,
we assure  compliance with Gibbs criteria for equilibrium.
Just
as in the case of the deconfinement phase transition, we find that the two
phases in equilibrium are oppositely charged, though in sum, neutral, as ought
to be so for stellar material. Consequently, the total energy, including
Coulomb and surface energies, is minimized by a lattice arrangement of the
rare phase immersed in the dominant \cite{glen91:d,glen91:a}.  The difference
between normal and kaon condensed phases is especially illustrated by the fact
that not only are kaons present in the latter, but the nucleon masses are
strongly modified from their values in the normal phase or in vacuum. Even
in the  spatial
regions of the mixed phase occupied by the normal or condensed
phase, the nucleons have different masses according to the phase
\cite{glen98:f}.  The high degree of inhomogeniety in the mixed phase
occupied as it is by a
lattice structure, the localization of opposite charge in the phase occupying
the lattice sites as compared with that of the background phase and the very
different nucleon effective masses in the two phases, will likely affect the
transport and superfluid properties of neutron stars.


The paper is organized as follows: in sec.\ \ref{sec:model} we introduce the
model Lagrangian which is based on the relativistic mean-field model for the
nucleon-nucleon interactions and a kaon-nucleon interaction motivated from
one-boson exchange. We discuss the equation of state and especially the 
difference between a Maxwell construction and Gibbs condition as well as the
properties of the mixed phase in
Sec.\ \ref{sec:EOS}. Consequences for the stellar properties are derived in
sec.\ \ref{sec:stellar}, both for global features as well as for the
resulting geometrical structures inside neutron stars. Our results are
summarized in sec.\ \ref{sec:summary}.


\section{Relativistic Mean--Field Model with Kaons}
\label{sec:model}

In the approach presented here, we use a relativistic nuclear field theory
solved in the mean-field approximation.  The interaction between baryons is
mediated by the exchange of scalar and vector mesons.  This picture is
consistently extended to include the kaons.  The model is similar to the
one used for describing the properties of the H dibaryon in nuclear matter
which is known to be thermodynamically consistent \cite{Fae97a}.  The coupling
schemes applied for the kaon are in analogy to the one we used for the H
dibaryon \cite{GS98}.

We start by summarizing briefly the relativistic mean-field model for
nucleons. 
The Lagrangian is given by
\begin{eqnarray}
{\cal L}_N &=& 
  \bar \Psi_N \left(i \gamma_\mu \partial^\mu - m_N 
  + g_{\sigma N} \sigma - g_{\omega N} \gamma_\mu V_\mu 
  - g_{\rho N} \vec{\tau}_N \vec{R}_\mu \right) \Psi_N
\cr &&
+ \frac{1}{2} \partial_\mu \sigma \partial^\mu \sigma 
- \frac{1}{2} m_\sigma^2 \sigma^2 - U(\sigma) 
\cr &&
- \frac{1}{4} V_{\mu\nu} V^{\mu\nu} 
+ \frac{1}{2} m_\omega^2 V_\mu V^\mu 
- \frac{1}{4} \vec{R}_{\mu\nu} \vec{R}^{\mu\nu} 
+ \frac{1}{2} m_\rho^2 \vec{R}_\mu \vec{R}^\mu 
\quad ,
\end{eqnarray} 
where $ V_{\mu\nu} \equiv \partial_\mu V_\nu - \partial_\nu
V_\mu $.  The scalar meson is denoted by $\sigma$, the vector meson $
{\mbox{\mbox{\boldmath$\omega$}}}$ by $V_\mu$ and the iso-vector $
{\mbox{\mbox{\boldmath$\rho$}}}$ meson by $R_\mu$.  The scalar
self-interactions $U(\sigma)$ are taken to be \cite{Bog77}
\begin{equation}
U(\sigma) = \frac{1}{3} b\, m_N (g_{\sigma N} \sigma)^3
          + \frac{1}{4} c\, (g_{\sigma N} \sigma)^4
\label{eq:selfint}
\quad .
\end{equation} 
The model parameters can be algebraically determined by five
bulk properties of nuclear matter \cite{book}.  Here, for illustrative
purposes, we choose one of
the parameter sets used in \cite{GM91} with the nuclear
matter properties: $E/A=-16.3$ MeV, $\rho_0=0.153$ fm$^{-3}$, $a_{{\rm sym}}=
32.5$ MeV, $K=240$ MeV, and $m^*/m=0.78$.  Other parameterizations will not
change the overall feature of kaon condensation as discussed in this paper.

Now we discuss the inclusion of the kaon-nucleon interaction terms.  There are
two main schemes for including effects of kaon condensation in neutron star
matter. One uses terms derived from chiral perturbation theory---the other
couples the kaon to meson fields. We choose to take the latter approach so that
nucleon and kaon interactions are treated on the same footing as pointed out
above.  The kaon is then coupled to the meson fields using minimal coupling
\begin{equation}
{\cal L}_K = {\cal D}_\mu^* K^* {\cal D}^\mu K - {m^*_K}^2 K^*K
\label{eq:lagK}
\end{equation}
where the vector fields are coupled via the standard form
\begin{equation}
{\cal D}_\mu = \partial_\mu + i g_{\omega K} V_\mu + 
i g_{\rho K} \vec{\tau}_K \vec{R}_\mu 
\label{eq:min_coupl}
\quad .
\end{equation}
Then the vector fields are coupled to a conserved current which is consistent
with Ward identities. The form (\ref{eq:min_coupl}) results in
another coupling term in the Lagrangian 
(\ref{eq:lagK}) of the form 
\begin{equation}
2 g_{\omega K}^2 V_\mu V^\mu K^* K
\end{equation}
in addition to the standard Yukawa coupling term
which gives a nonlinear dependence of the kaon optical potential with density.

The scalar field is coupled to the kaon by analogy to the minimal coupling
scheme of the vector fields
\begin{equation}
m^*_K = m_K - g_{\sigma K} \sigma \quad .
\label{eq:scoupl}
\end{equation}
In addition to the standard linear Yukawa coupling term, it gives also 
a quadratic coupling term to the scalar field in the Lagrangian of the form
\begin{equation}
(g_{\sigma K} \sigma)^2 K^* K
\quad .
\end{equation}
This term is small compared to the linear Yukawa coupling term as it is 
suppressed by $g_{\sigma K}/(2m_K)$. 
Nevertheless, it will simplify the equations
of motion considerably as we will show in the following.

The equation of motion for the kaon can be written as
\begin{equation}
\left[{\cal D}_{\mu} {\cal D}^\mu 
 + {m^*_K}^2 \right] K = 0 
\quad .
\end{equation}
The poles of the kaon propagator can then be determined by
\begin{equation}
-\omega_K^2 + m_K^2 + k^2 + \Pi_K (\omega_K,\vec{k},\rho) = 0
\end{equation}
where the K$^-$  self-energy in matter 
(the space components of the vector field vanish $V_i=\vec{R}_i=0$)
is given by
\begin{eqnarray}
\Pi_K (\omega,\vec{k},\rho) &=&
- 2\omega_K \left(g_{\omega K}V_0 + g_{\rho K}\vec{\tau}_K \vec{R}_0\right)
- \left(g_{\omega K}V_0 + g_{\rho K}\vec{\tau}_K \vec{R}_0\right)^2
\cr &&
- 2m_K g_{\sigma K} \sigma + (g_{\sigma K} \sigma)^2
\end{eqnarray}
and depends on the in-medium kaon energy $\omega_K$.
It is straightforward to derive the dispersion relation 
for s-wave condensation (i.e.\ for $\vec{k}=0$) for the K$^-$
\begin{equation}
\omega_K = m_K - g_{\sigma K}\sigma -
g_{\omega K} V_0 - g_{\rho K} R_{0,0}
\label{eq:disp}
\end{equation}
which is linear in the meson fields.
There appear additional source terms in the equation of motion for the meson
fields if a kaon condensate is present
\begin{eqnarray}
m_\sigma^2 \sigma &=& - b\,m_N (g_{\sigma N} \sigma)^2 
- c\,(g_{\sigma N} \sigma)^3 +
g_{\sigma N} \rho_s + 2 g_{\sigma K}  m^*_K K^* K \cr
m_\omega^2 V_0 &=&
g_{\omega N} (\rho_p + \rho_n) - 2 g_{\omega K} (\omega_K + g_{\omega K} V_0 +
g_{\rho K} R_{0,0} ) K^* K \cr
m_\rho^2 R_{0,0} &=&
g_{\rho N} \left(\rho_p -\rho_n\right) - 2 g_{\rho K} 
(\omega_K + g_{\omega K} V_0 + g_{\rho K} R_{0,0} ) K^* K
\quad .
\end{eqnarray}
Note that the equation of motion for nucleons are unchanged.
The conserved current associated with the kaons is derived by using
\begin{eqnarray}
J_\mu^K &=& i \left(K^* \frac{\partial \cal L}{\partial^\mu K^*}
- \frac{\partial \cal L}{\partial^\mu K} K \right) \cr
&=& K^* i\partial_\mu K -(i\partial_\mu K^*)K 
- 2g_{\omega K} V_\mu K^* K - 2g_{\rho K} \vec{\tau}_K \vec{R}_\mu K^* K 
\quad .
\end{eqnarray}
In the mean-field approximation, the K$^-$ density is given by
\begin{eqnarray}
\rho_K=-J_0^K = 
2\left(\omega_K + g_{\omega K} V_0 + g_{\rho K} R_{0,0}\right) K^* K \quad .
\end{eqnarray}
For s-wave condensation we can use the dispersion relation 
(\ref{eq:disp}) to get an expression for the scalar density of the kaon
\begin{equation}
2 m^*_K K^* K = 2 (\omega_K + g_{\omega K} V_0 
+ g_{\rho K} R_{0,0} ) K^* K = \rho_K
\end{equation}
which comes out to be the same as the vector density. 
This relation holds only for $\vec{k}=0$ which is the case for cold
neutron star matter and s-wave condensation.
It is a result of our choice of the scalar coupling scheme (\ref{eq:scoupl}).
For the negatively charged kaon the equation of motion are 
then simplified to
\begin{eqnarray}
m_\sigma^2 \sigma + b\,m_N (g_{\sigma N} \sigma)^2 
+ c\,(g_{\sigma N} \sigma)^3 &=&
g_{\sigma N} \rho_s + g_{\sigma K} \rho_K \cr
m_\omega^2 V_0 &=&
g_{\omega N} (\rho_p + \rho_n) - g_{\omega K} \rho_K \cr
m_\rho^2 R_{0,0} &=&
g_{\rho N} \left(\rho_p - \rho_n\right) - g_{\rho K} \rho_K
\quad .
\label{eq:eomK}
\end{eqnarray}
The total energy density is given by
\begin{equation}
\epsilon = \epsilon_N  + \epsilon_K  
\end{equation}
and has a contribution from the kaon condensate.
The nucleon part consists of  the standard terms (cf. Ref. \cite{book})
\begin{eqnarray}
\epsilon_N &=& 
\frac{1}{2}m_\sigma^2 \sigma^2
+ \frac{b}{3}\,m_N (g_{\sigma N} \sigma)^3 
+ \frac{c}{4}\,(g_{\sigma N} \sigma)^4
+ \frac{1}{2}m_\omega^2 V_0^2 
+ \frac{1}{2}m_\rho^2 R_{0,0}^2
\cr && 
+ \sum_{i=N,l} \frac{\nu_i}{(2\pi^3)} \int_0^{k_F^i} \!\!\! d^3 k
\sqrt{k^2 + {m^*_i}^2}
\end{eqnarray} 
The sum is over nucleons and leptons.
In principle it could extend over baryons of the octet, but we neglect 
the higher members in the present study. 
The kaon contribution to the energy density reads
\begin{equation}
\epsilon_K = 2 {m^*_K}^2 K^*K = m^{*}_{K} \rho_K
\quad.
\label{eq:ekaon}
\end{equation}
The kaon does not contribute directly to the pressure 
as it is a (s-wave) Bose condensate so that the total pressure
\begin{eqnarray}
p &=& 
- \frac{1}{2}m_\sigma^2 \sigma^2
- \frac{b}{3}\,m_N (g_{\sigma N} \sigma)^3 
- \frac{c}{4}\,(g_{\sigma N} \sigma)^4
+ \frac{1}{2}m_\omega^2 V_0^2 
+ \frac{1}{2}m_\rho^2 R_{0,0}^2
\cr && 
+ \sum_{i=N,l} \frac{\nu_i}{(2\pi^3)} \int_0^{k_F^i} \!\!\! d^3 k
\frac{k^2}{\sqrt{k^2 + {m^*_i}^2}}
\end{eqnarray} 
is just the familiar expression known from relativistic mean
field theory for nucleons and leptons only.  Hence, the equation of state will
be considerably softened if the kaon condensate is present. The pressure is
modified only indirectly through the change of the meson fields by the
additional kaon source terms which enter into the equations of motion
(\ref{eq:eomK}).  The total charge is then
\begin{eqnarray}
q_N &=& \rho_p - \rho_e - \rho_\mu \\
q_K &=& \rho_p - \rho_e - \rho_\mu - \rho_K
\end{eqnarray} 
in the normal and in the kaon condensed phase, respectively.

The above relations do not fix the amplitude of the kaon condensate $K^*K$.
The charged kaon amplitude is zero unless the condition
\begin{equation}
\omega_K = \mu_{K^-} = \mu_e
\label{eq:cond}
\end{equation}
can be fulfilled.  Generally the electrochemical potential increases
as the baryon density increases, since this will usually mean that
the proton density increases. Moreover, the K$^-$ effective mass in the
medium decreases  with increasing density. Therefore at some density
the above threshold condition may be fulfilled. Since all kaons can
condense in the lowest energy state, they become energetically more favorable
than electrons as the neutralizing agent of positive
charge. With further increase of
density and decrease in kaon energy $\omega_K$, the electrochemical potential
will decrease and the electron population will decrease.

For the isospin partner of the K$^-$, the K$^0$, the condition for condensation
to happen reads $\omega(K^0) = 0$. 
Hence, if there is no isovector potential for
the kaons, the K$^0$ can only appear
after charged kaons have appeared and the electrochemical potential 
hits zero. This seems quite unlikely and, to our knowledge, was therefore
completely ignored in previous works. Nevertheless, the isospin potential of
the nucleons shift the energy of the K$^0$ below the one of the K$^-$ in
neutron-rich matter so that
\begin{equation}
\omega(K^0) = \omega(K^-) + 2 g_{\rho K} R_{0,0} =
\omega(K^-) - 2 \frac{g^2_{\rho K}}{m_\rho^2} \left(\rho_n-\rho_p\right)
\quad .
\end{equation}
A strong isovector potential is
supported by coupled channel calculations for the K$^-$ \cite{Waas96b} which
shifts the effective energy of the K$^-$ up by approximately 100 MeV at a
density of $\rho=3\rho_0$. This would imply that the effective energy for the
K$^0$ is about 200 MeV lower than the one for the K$^-$ in neutron matter at
that density!
In our calculations, we find indeed that K$^0$ condensation can happen if the
isovector coupling constant is chosen as strong as the nucleon one (see eq.\
(\ref{eq:iso_coupl}) in the following). 
For the sake of simplicity we ignore it in the following but note
that it is clear from our discussion that 
K$^0$ condensation should be taken into account in a more realistic
calculation. 

The Lagrangian for the kaons (\ref{eq:lagK}) describes 
the kaon-nucleon interaction
as well as the kaon-kaon interaction. 
The K$^-$ in a nuclear medium is certainly a coupled
channel problem due to the opening of the $\Sigma\pi$, $\Lambda\pi$ channels
and can not be treated on the mean-field level. Coupled channel
calculations at finite density, first done by Koch \cite{Koch94}, 
yield an attractive potential
for the K$^-$ at normal nuclear density of about 
$U_{K^-} (\rho_0) = -100$ MeV.
Waas et al.\ find a value of 
$U_{K^-} (\rho_0) = -120$ MeV \cite{Waas97}.
 Kaonic data support the
 conclusion that there is a highly
 attractive kaon optical potential in dense nuclear matter \cite{Fried94}.
Because the kaon is a boson it does not add directly to the pressure;
it
forms a Bose condensate in the s-wave with zero momentum \cite{KN86}.  This is
contrary to pion condensation which condenses in a p-wave with a finite
momentum.  A selfconsistent treatment of the in-medium self energy of the pion
prevents pion condensation \cite{Dick83}.  A coupled channel calculation
including the modified self-energy of the kaon has been studied in
\cite{Lutz98} and it was found that the kaon still sees an attractive potential
at high density.

On the mean-field level considered here, 
the three kaon coupling constants, $g_{\sigma K}$, $g_{\omega K}$, 
and $g_{\rho K}$ can be fixed to 
kaon-nucleon scattering lengths. The in-medium 
potentials for the K$^-$ are given by G-parity, i.e.\ by switching the sign of
the vector potential. This gives similar results for the K$^-$ optical
potential compared to the coupled channel calculations \cite{SMB97}.
We choose to couple the vector fields according to the simple quark
and isospin counting rule
\begin{equation}
g_{\omega K} = \frac{1}{3} g_{\omega N} \quad \mbox{ and } \quad
g_{\rho K} = g_{\rho N} 
\quad.
\label{eq:iso_coupl}
\end{equation}
The scalar coupling constant is fixed to the optical potential of the K$^-$ at
$\rho_0$:
\begin{equation}
U_K (\rho_0) =  -g_{\sigma K} \sigma( \rho_0) - g_{\omega K} V_0 (\rho_0)
\quad 
\end{equation}
The kaon potential is fixed at normal nuclear density
and varies as a function of the density $\rho$.

We solve the equations of motions in three different ways corresponding to the
three possible solutions: 1) for pure nuclear matter without kaons,
2) for pure kaon condensed matter,
3) for the mixed phase. The latter one is found by solving for solution 1) and
2) separately and scanning through the
electrochemical potential until the pressures in the two phases 
for the same chemical potentials are equal. Initially, values for the meson
fields are taken randomly.
The solution found at a certain baryochemical potential is then used for the
next step. We compare then the pressure of the three solutions and take the one
with the highest pressure. It turns out that this procedure ensures
automatically that the solution for the mixed phase gives a (thermodynamically
consistent) volume fraction between zero and one.


\section{Equation of State with a Kaon Condensate}
\label{sec:EOS}

In the following we discuss the equation of state including kaons 
emphasizing the difference
between the hitherto applied Maxwell construction and the thermodynamic
consistent Gibbs condition. Then we present our results for the two phases in
the mixed phase. 

\subsection{Maxwell versus Gibbs}

The standard thermodynamic rule for two phases in thermodynamical equilibrium
is given by the Gibbs condition 
\begin{equation}
p^{\rm I} = p^{\rm II} \,,\quad \mu_i^{\rm I} = \mu_i^{\rm II}\,, \quad
T^I=T^{II}
\label{eq:gibbscond}
\end{equation} 
which simply states that the two phases are in mechanical,
chemical and thermal equilibrium. This is basic thermodynamics and can be found
in textbooks.  For the special case of only {\em one} chemical potential, the
resulting equation $p^{\rm I}(\mu,T)=p^{\rm II}(\mu,T)$ has a unique solution
for $\mu$.  It is often found by use of a Maxwell construction in one form or
another. For example, the common tangent method, is based on the fact that
$\mu=d\epsilon/d\rho =dE/dN$. Write the equation of state in the form
$\epsilon=\epsilon(\rho)$.  The segment of the common tangent,
$\epsilon=-p_0+\mu_0 \rho$ touching the equation of state once in each phase
describes the mixed phase with common and constant values of $p_0$ and $\mu_0$,
independent of the proportion of the two phases. Clearly, the Maxwell
construction can assure that only a single chemical potential is common to both
phases.  

However, neutron star matter has two chemical potentials, $\mu_B$ and
$\mu_e$, each of which must be equal in the two
phases to assure equilibrium. Hence, a
Maxwell construction {\sl cannot} be used as it will produce a discontinuity
in one of the chemical potentials and will describe an unstable state---one
for which there is a potential difference at the boundary between phases.
 This
general fact concerning phase transitions with more than one conserved charge
inside neutron stars was realized only a few years ago 
\cite{glen91:d,glen91:a}.  It was
shown how to assure 
equilibrium in substances of an arbitrary number of conserved charges, 
that conservation laws cannot be locally imposed but only globally over the
entire region of mixed phase, how internal forces redistribute conserved
charges between equilibrium phases so as to minimize the energy,
and how, in the case that the electric charge is among the conserved charges,
a Coulomb lattice will be formed.

Conservation laws and Gibbs conditions can be satisfied simultaneously
for substances of more than one conserved charge by applying the
conservation law(s) only in a {\sl global} rather than a {\sl local}
sense 
\cite{glen91:d,glen91:a}.
Thus for neutron star matter which has two conserved
charges, the Gibbs conditions and conservation of electric charge
read
\begin{eqnarray}
p_N(\mu_B, \mu_e) =p_K(\mu_B, \mu_e)  \label{pres}\\
q_{{\rm total}}= (1-\chi)q_N(\mu_B, \mu_e) +\chi q_K(\mu_B, \mu_e) = 0\,.
\label{total}
\end{eqnarray}
where $q$ denotes charge density of the corresponding phase.
This pair of equations can be solved for $\mu_B, {\rm ~and~} \mu_e$
for any volume
proportion of kaon phase $\chi$ in the interval (0,1).
Therefore the chemical potentials are functions of proportion $\chi$
and therefore also are all other properties of the two phases, including
the common pressure. 
This is only a mathematical proof that in general, 
properties will vary as the proportion. 
{\sl Why} and how they vary
depends on how the internal driving forces can exploit
the degrees of freedom (one less than the number of independent 
chemical potentials) so as to minimize the total energy \cite{glen91:a}.

The total baryon density in the mixed phase corresponding to the 
solution of the above pair of equations for charge neutral matter in
phase equilibrium is given as a function of $\chi$ by
\begin{equation}
n_{{\rm total}}= (1-\chi)n_N(\mu_B, \mu_e) +\chi n_K(\mu_B, \mu_e) 
\,,
\end{equation}
where $n$ denotes baryon number density.
A similar equation holds for the energy density.
It will be noted that the pressure equality 
(\ref{pres}) 
cannot be solved simultaneously with conditions of {\sl local} charge
neutrality, $q_N(\mu_B, \mu_e)=0,~~ q_K(\mu_B, \mu_e) = 0$, since three
conditions must be satisfied with only two variables. 

\begin{figure}[htbp]
\begin{center}
\leavevmode    
{\psfig{figure=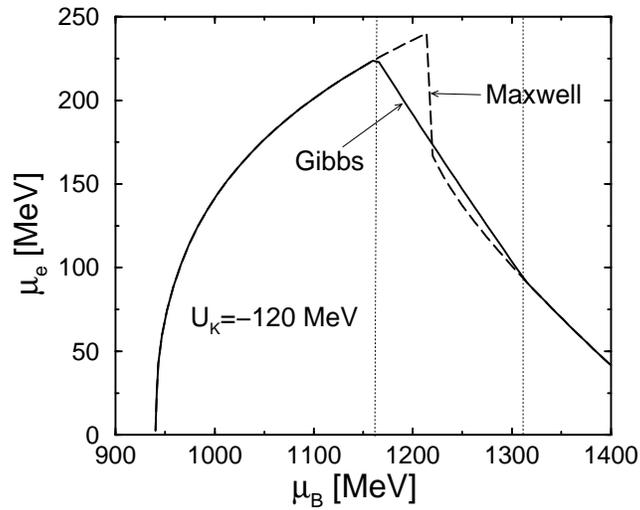,height=0.35\textheight}\vspace*{-1cm}}
\parbox[t]{4.6in}{\caption{\label{fig:chempot}
The two chemical potentials, the electrochemical
potential and the baryochemical potential for the case $U_K(\rho_0)=-120$ MeV
using the Gibbs condition (solid line) and a Maxwell construction (dashed
line). The  large electric potential difference
that occurs
for the Maxwell
construction  gives rise to an instability.}}\end{center}  
\end{figure}

Figure~\ref{fig:chempot} shows the
behavior of the chemical potentials using Gibbs and Maxwell construction for
comparison. 
The vertical dotted lines indicate the region of the mixed phase
when using the Gibbs condition implemented
for charge neutrality as described above. 
The electrochemical potential increases
in the pure hadronic phase as the density of neutrons and protons 
increase. However, at the critical density for kaon condensation
the  electrochemical potential becomes a decreasing function of density
as kaons replace electrons in their role of neutralizing the charge on 
protons. We note that when the conservation of electric charge is imposed
as a global constraint, as described above and in Ref. \cite{glen91:a},
the electrochemical potential is continuous, in contrast to the case 
of the Maxwell construction.
In the Maxwell
construction, the electrochemical potential drops from $\mu_e=240$ MeV 
to $\mu_e=167$ MeV at the phase boundary
resulting in a huge difference in the Fermi energy of the leptons 
between the two phases. There needs to be an additional force to prevent the
electrons moving from the phase with the higher chemical potential to the
other which is completely absent in a bulk treatment. 

\begin{figure}[htbp]
\begin{center}
\leavevmode    
{\psfig{figure=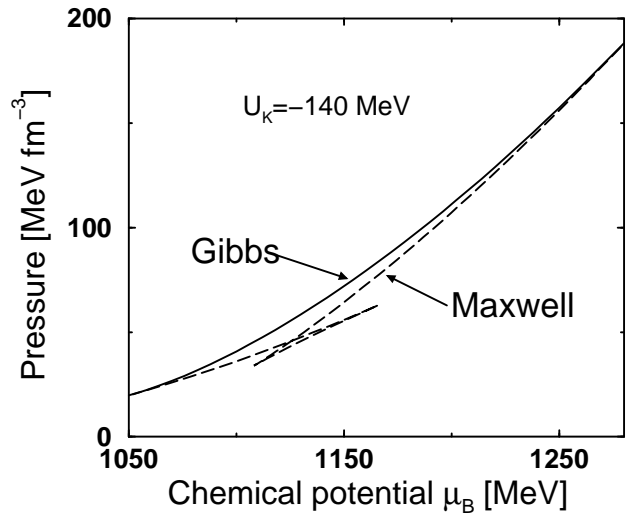,height=0.35\textheight}\vspace*{-1cm}}
\parbox[t]{4.6in}{\caption{\label{fig:pressure}
The pressure versus the baryochemical potential for a Maxwell construction
(dashed line) compared to the Gibbs condition (solid line). The Gibbs
condition is thermodynamically more stable.}}
\end{center}
\end{figure}

A Maxwell construction is often implemented
by looking at the thermodynamical
potential of interest, here the pressure, as a function of the chemical
potential as depicted in Fig.\ \ref{fig:pressure}. The crossing of the curve is
the point of equal pressure at the same baryochemical potential. 
Using the Gibbs conditions (\ref{eq:gibbscond}), the solid curve results which
has 
always a higher pressure compared to the Maxwell construction being the
thermodynamically favored one. The change in the slope at the crossover of the
Maxwell construction is smeared out. The pressure difference between the two
cases depends on the  equation of state and the optical potential of the
kaon. In addition, it is also sensitive to finite size corrections. Here, we
discuss only bulk matter. Coulomb energy and surface energy will reduce the
pressure in the mixed phase. For the mixed phase of normal nuclear matter and
nuclei in the crust of the neutron star, this correction is on
the order of 10 MeV/fm$^{1}$. It depends on the surface tension
which is unknown for a kaon condensed phase immersed in dense nuclear matter
and will shift the curve for the
Gibbs condition case to slightly lower values.  But because the
sum of  Coulomb energy and surface energy vanishes at the boundaries of the
mixed phase (cf. eq. (2) in Ref.
\cite{glen95:c}), the boundaries are unaffected. 
We will discuss the geometric features when kaons
are condensing  in more detail later.

\begin{figure}[htbp]
\begin{center}
\leavevmode    
{\psfig{figure=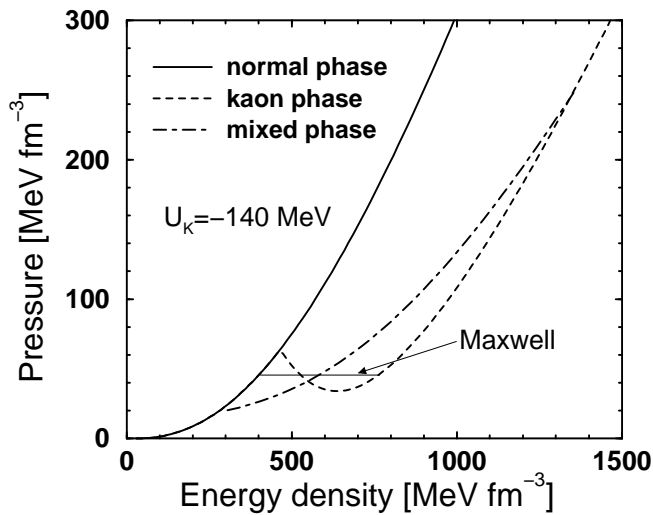,height=0.35\textheight}\vspace*{-1cm}}
\parbox[t]{4.6in}{\caption{\label{fig:eos_maxwell}
The equation of state for a pure nuclear matter (solid line), pure kaon matter
(dotted line) for $U_K(\rho_0)=-140$ MeV. 
The Maxwell construction is shown by the horizontal line, the
Gibbs solution by the dashed-dotted line.}}\end{center}
\end{figure}

The differences between the two descriptions, Maxwell and Gibbs, are most
striking for the relevant observable for neutron star calculations: 
the equation of state as plotted in Fig.\ \ref{fig:eos_maxwell}. 
The solid line shows the equation of state for
the normal hadronic phase of neutron star matter, the dotted line
the one for pure kaon condensed matter. The Maxwell construction results in a
region of constant pressure (solid horizontal line) connecting the two
different equation of states. Applying the Gibbs condition causes two major
differences compared to the Maxwell construction. First, the region of constant
pressure vanishes and there is a continuous increase of the pressure. 
Second, the density range of the mixed phase is much wider, it starts at a
lower density and ends at a much higher density. Hence, the mixed phase can
well be the dominant portion of a neutron star.

\begin{figure}[htbp]
\begin{center}
\leavevmode    
{\psfig{figure=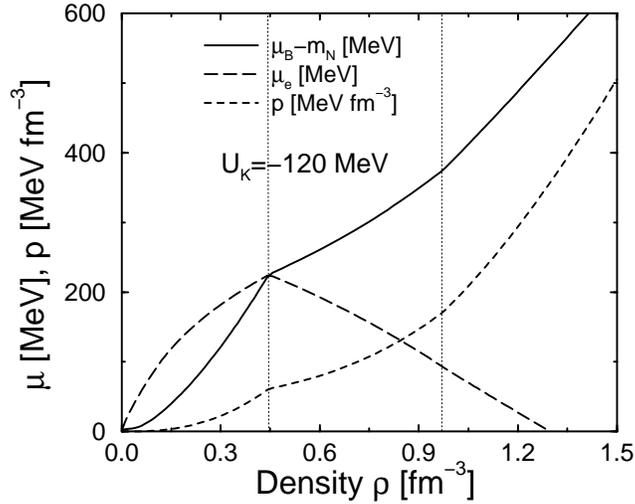,height=0.35\textheight}\vspace*{-1cm}}
\parbox[t]{4.6in}{\caption{\label{fig:mu120}
The pressure, the electrochemical potential, and the baryochemical potential
are plotted over the mixed phase region using the Gibbs
condition.}}\end{center} 
\end{figure}

The behavior of the thermodynamic potential, the pressure, and the two chemical
potentials over the mixed phase region using the Gibbs condition 
is summarized in Fig.\ \ref{fig:mu120}. 
The baryochemical potential as well as the pressure are continuously rising
with density. The electrochemical potential  increases until the mixed phase
starts, then it is continuously decreasing with density. There is now no jump
in any of these observables and none stays constant over the mixed phase
region. 

\begin{figure}[htbp]
\begin{center}
\leavevmode    
{\psfig{figure=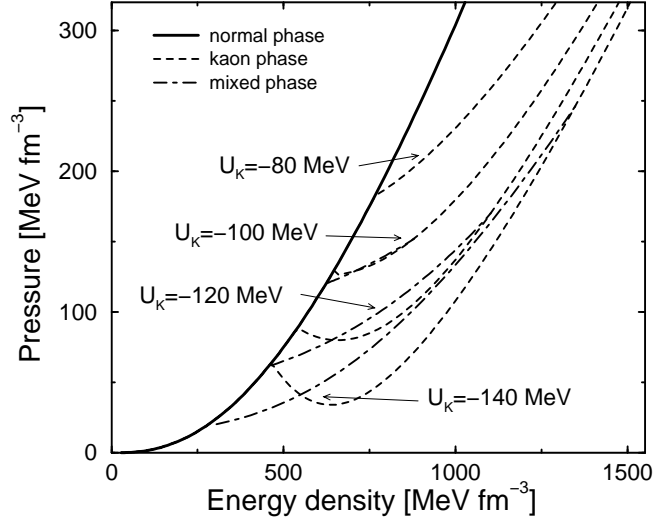,height=0.35\textheight}\vspace*{-1cm}}
\parbox[t]{4.6in}{\caption{\label{fig:eos_all}
The equation of state for various choices of the optical potential of the
kaon.}}\end{center} 
\end{figure}

The form of the equation of state depends sensitively on the chosen optical
potential of the kaon. Figure \ref{fig:eos_all} shows the equations of state
for optical potentials of the kaon at normal nuclear matter density between
--80 and --140 MeV.  For $U_K(\rho_0)=-80$ MeV, there is no mixed phase and the
phase transition is of second order. For a deeper optical potential, a mixed
phase appears as plotted in dashed-dotted lines. The deeper the optical
potential of 
the kaon is, the lower is the density the mixed phase starts and the wider is
the range of the mixed phase. The equation of state is considerably softened by
the presence of the kaon condensate.  The critical density for the onset of the
kaon condensed phase is summarized for various kaon optical potentials in
Table~\ref{tab:crit_dens}. For the cases $U_K(\rho_0)=-80$ MeV and $-90$ MeV
the phase transition is of second order.

\begin{table}[htbp]
\begin{center}
\begin{tabular}{|c|c|c|c|c|c|c|c|c|}
\hline\hline
$U_K(\rho_0)$ (MeV) & --80 & --90 & --100 & --110 & --120 & --130 & --140 &
--150 \cr
\hline
$\rho_c/\rho_0$ & 4.5 & 4.2 & 3.8 & 3.4 & 2.9 & 2.4 & 1.8 & 1.2
\cr \hline\hline
\end{tabular}
\parbox[t]{4.6in}{\caption{The critical density for the appearance of the kaon
condensed phase for different kaon optical potentials.}}
\end{center}
\label{tab:crit_dens}
\end{table}

\begin{figure}[htbp]
\begin{center}
\leavevmode    
{\psfig{figure=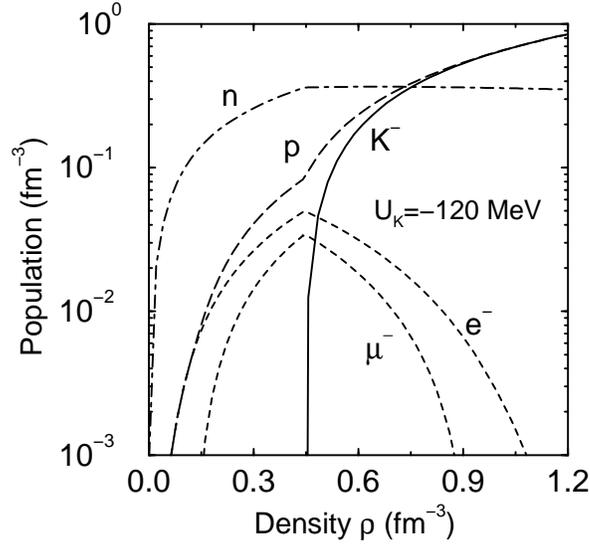,height=0.35\textheight}\vspace*{-1cm}}
\parbox[t]{4.6in}{\caption{\label{fig:pop_u120}
The population as a function of the nucleon density. The neutron density stays
nearly
constants once kaon condensation appears.}}\end{center}
\end{figure}

Fig.\ \ref{fig:pop_u120} shows the populations of the nucleons, leptons and
kaons for the case $U_K(\rho_0)=-120$ MeV. The remarkably feature is the
'frozen' neutron density once kaons start to condense. As it is more favorable
to produce kaons in association with protons, the neutron density just stays
(nearly) constant over the whole density range shown starting with the critical
density. The lepton populations decrease as the K$^-$ appears as the
new neutralizing agent.

The neutron density seems to be frozen once kaons appear in the system as
viewed
on a logarithmical scale,  but it  actually varies slowly,
going up and then down slightly with density
at the order of a few percent. Note that the overall neutron density in the
mixed phase is the sum of the two contributions from the normal and the kaon
phase which makes it even more puzzling. The neutron population does even not
change when the pure kaon phase is reached. Nevertheless, comparison with
previous work on kaon condensation also indicates that the neutron population
does not change very much once kaon condensation sets in. From Tables 3 and 4
in Ref.\ \cite{Brown94} one can read off the neutron density and finds that it
changes at the level of a few percent up to moderate densities. This holds also
for the calculations done by Fujii et al.\ \cite{Fujii96}. The neutron
density actually decreases first after kaons have appeared then it rises again
for larger densities. The actual change in the neutron density is
less than 10\% up to a density of $\approx 5\rho_0$ after kaons have condensed
\cite{Hiro}. 

\begin{figure}[htbp]
\begin{center}
\leavevmode    
{\psfig{figure=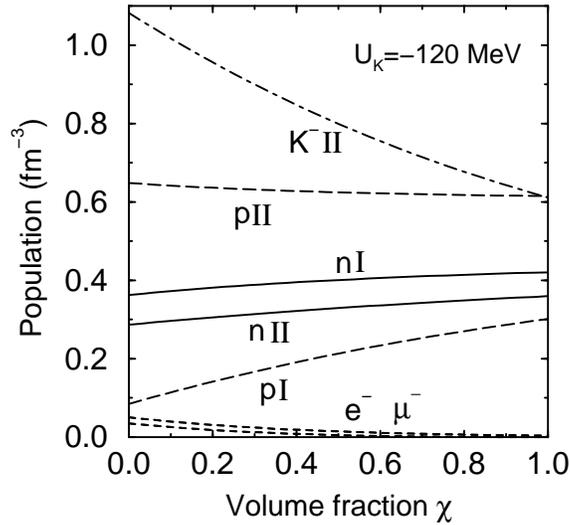,height=0.35\textheight}\vspace*{-1cm}}
\parbox[t]{4.6in}{\caption{\label{fig:populmix120}
The population as a function of volume fraction of kaon condensed
phase. Normal phase is denoted
by I and condensed phase by II. Note the finite charge density 
in the condensed phase (negative) and normal phase (positive) 
which vanish only on the boundaries $\chi=1 {\rm~or~}0$.}}\end{center}
\end{figure}

Figure \ref{fig:populmix120} shows the population in the mixed phase for the
two phases separately as a function of volume proportion $\chi$ of
condensed phase. The normal phase population is denoted as I, the kaon
phase population as II.  
For $\chi=0$ the proton population in the normal
phase is small and neutrality is achieved by a 
balance with the sum of the lepton populations. This corresponds to local
neutrality in the pure phase. However with a growing fraction
of condensed phase, charge neutrality is achieved more economically
between the two
phases in equilibrium as a global
constraint---the proton population increases to near equality with
neutrons  as the proportion of condensed phase increases,
while the lepton populations decrease to
the vanishing point. Isospin symmetry is thus closely achieved in the
normal phase. 
This behavior is expected and explained in 
Ref. \cite{glen91:a} as a general feature of  the action of the isospin driving
force toward symmetry in phase transitions of asymmetric nuclear matter.

The population behaviors in the condensed phase
are different.  We can understand this as follows: compare the energy of
neutron and of a  proton-K$^-$ pair, whose chemical potentials are the
same. They are respectively
\beqn
\mu_n=E_n &=&\sqrt{m^{\star 2} + k^{2}_{F,n} } 
+g_{\omega N} V_0 -g_{\rho N} R_{0,0}
\\
\mu_n=\omega_{K^-} + E_p &=& m_K - g_{\sigma K} \sigma -\frac{1}{3}g_{\omega N}
V_0 -g_{\rho N} R_{0,0}\nonumber\\
& & + \sqrt{m^{\star 2} +k^{2}_{F,p}} +g_{\omega N} V_0 +
g_{\rho N} R_{0,0}\nonumber\\
 &=& m_K - g_{\sigma K} \sigma + \sqrt{m^{\star 2} +k^{2}_{F,p}}
 +\frac{2}{3}g_{\omega N}
 V_0 \,.~~~~~~~~~
\eeqn
From these two expressions of $\mu_n$ we have,
\beqn 
 \sqrt{m^{\star 2} +k^{2}_{F,p}}=
 \sqrt{m^{\star 2} + k^{2}_{F,n} }+ \frac{1}{3}g_{\omega N}
 V_0  -g_{\rho N} R_{0,0}
 \,, \eeqn
 from which it is clear that $k_{F, p}>k_{F,n}$ when the sum of the last two
 terms is positive. Since $V_0$ is proportional to the density, whereas
 $R_{00}$ is proportional to the  difference in isospin densities of 
 proton and neutron, the sum will generally be positive, and is in the present
 case. That is, p-K$^-$ pairs are preferred to neutrons.
 However, the symmetry restoring term in the
 energy will prevent an uninhibited growth of protons compared to neutrons.

We have neglected  K$^0$ condensation (as has everyone else).
         It is clear that  eventually there will be a competition between
          p-K$^-$ pairs and  n-K$^0$ pairs. It appears that with increasing
          density, the condensed phase will tend toward symmetry in 
          neutrons and protons and similar density of K$^-$ and K$^0$.

\subsection{Mixed phase properties}

We will show in the following that the two phases in equilibrium in the mixed
phase have completely different properties. We will focus on the case
$U_K=-120$ MeV in this section.
 
\begin{figure}[htbp]
\begin{center}
\leavevmode    
{\psfig{figure=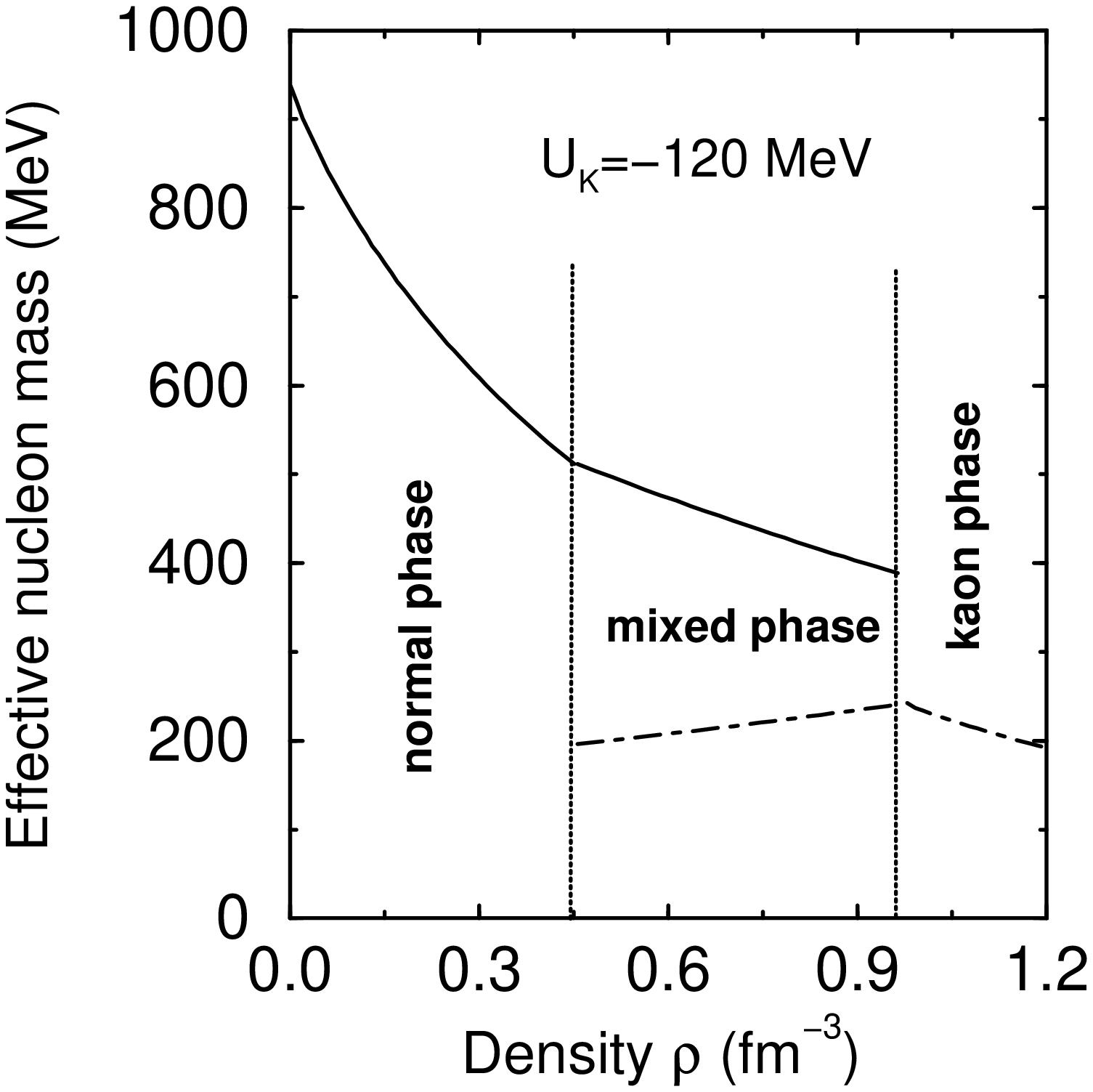,height=0.35\textheight}\vspace*{-1cm}}
\parbox[t]{4.6in}{\caption{\label{fig:effmass}
The effective nucleon mass as a function of the nucleon density. Shown by
vertical lines is the onset and offset of the mixed phase.}}\end{center}
\end{figure}

One striking question is, why should the nucleons not be the same in the two
phases and why can they not move freely between phase I and phase II.  The
answer is that when the nucleons are treated as dynamical particles they are
different in the two phases. Their interaction with the kaon field is what
causes the decrease of the kaon effective mass with increasing density. The
decrease in kaon mass ultimately leads to the condensation of kaons. The
interaction also changes the nature of a nucleon.  Figure \ref{fig:effmass}
illustrates the dynamical nature of the nucleon: its effective mass is shown as
a function of baryon density.  Up to $\rho=0.45$ fm$^{-3}$ there exists only
one solution---the pure nucleon phase. The effective mass decreases with
density from its vacuum value  to 0.78 of its vacuum value at the
saturation density of symmetric matter and further 
down to $m_N^*=510$ MeV at the end of the pure normal phase
($\sim 3 \rho_0$). In the mixed phase, a second solution appears
at a much lower effective nucleon mass of $m_N^*=196$ MeV. The second solution
is the nucleon effective mass in the kaon condensed phase fraction of the
mixed phase.  
The mixed phase ends at $\rho=0.97$ fm$^{-3}$ and only the second solution
continues, now changing slope and decreasing with density.  The nucleons have
different effective masses due to the different mean-fields in the two phases
in equilibrium.  Hence, the nucleons cannot move freely between the two phases
and a phase boundary can develop.  In ref.\ \cite{Thorsson94} the nucleons in
the two phases were treated only implicitly through a phenomenological equation
of state.  They did not appear as dynamical degrees of freedom, so the two
solutions could not be found.

\begin{figure}[htbp]
\begin{center}
\leavevmode    
{\psfig{figure=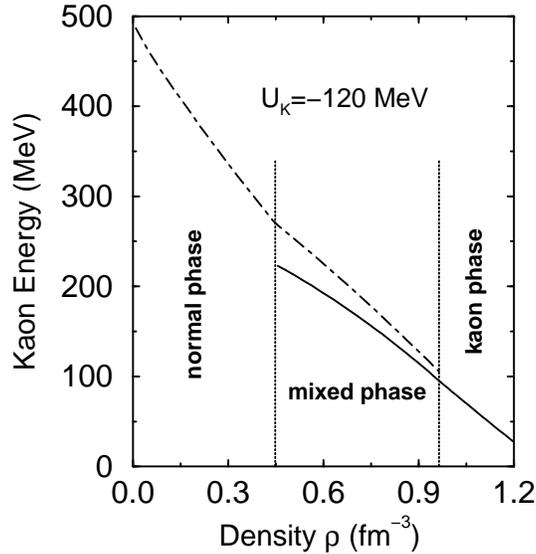,height=0.35\textheight}\vspace*{-1cm}}
\parbox[t]{4.6in}{\caption{\label{fig:kaon_energy}
The effective kaon energy versus the nucleon density.}}\end{center}
\end{figure}

Figure \ref{fig:kaon_energy} depicts the analogue to Fig.\ \ref{fig:effmass}
for the effective energy of the kaon. Note that the kaon is only a test particle
in the nucleon (normal) phase and appears only physically in the condensed
phase.
The effective kaon energy decreases with density due to the
attractive vector interaction with nucleons, but kaons do not appear in the
medium until the threshold condition discussed above is satisfied. However, we
can trace the energy of a test kaon in the medium and it is shown in Fig.\ 
\ref{fig:kaon_energy} as dashed-dotted line.  
Its energy as a test particle is also shown in the regions
of the mixed phase that are occupied by the normal phase. When the kaon energy
sinks to a value satisfying the threshold condition, kaons begin to appear, but
because the phase transition is first order they first appear
in a small fraction of the total volume which is the kaon condensed phase in
equilibrium with the normal phase.  The energy of these medium modified kaons
is less than that of a test kaon in the normal phase.  The two energies are
shown in the figure.

\begin{figure}[htbp]
\begin{center}
\leavevmode    
{\psfig{figure=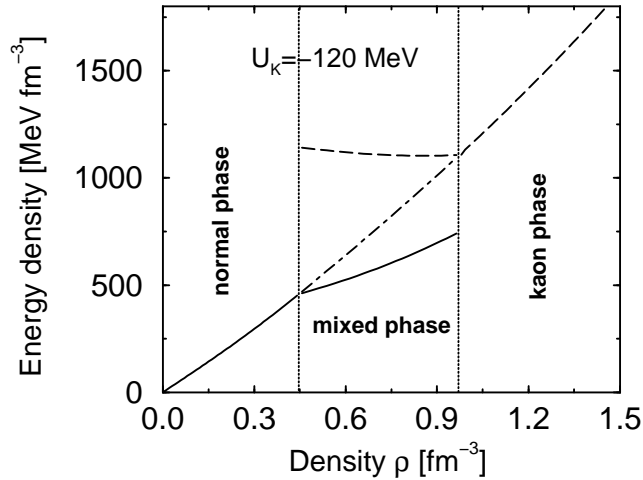,height=0.35\textheight}\vspace*{-1cm}}
\parbox[t]{4.6in}{\caption{\label{fig:enrg120}
The energy density  of normal phase (solid) and kaon condensed phase
(dashed). The total energy is the volume weighted sum (dash-dotted).}}
\end{center}
\end{figure}

The most pronounced differences between the two phases is in the energy density
and the charge density. As can be read from Fig.\ \ref{fig:enrg120}, the energy
density of the nucleon phase (solid line) at the onset of the mixed phase is
$\epsilon=460$ MeV fm$^{-3}$ while it amounts to $\epsilon=1140$ MeV fm$^{-3}$
for the kaon condensed phase (dashed line). The dashed-dotted line is the sum
of the energy density of the two phase according to their volume fraction
$\chi$:
\begin{equation}
\epsilon = (1-\chi) \epsilon_N(\chi) + \chi \epsilon_K(\chi)
\end{equation} 
and is continuously growing with density but {\sl not} linearly,
as is the case in the Maxwell construction 
($\epsilon_K(\chi)$ denotes the total energy density of the kaon phase and
should be distinguished from eq.\ (\ref{eq:ekaon})).
The  non-constant pressure  is of course
associated with the 
non-linearity.




\section{Stellar Properties}
\label{sec:stellar}

\subsection{Large scale features}
\label{sec:gross}

\begin{figure}[tbh]
\vspace{-.4in}
\begin{center}
\leavevmode
\centerline{ \hbox{
\psfig{figure=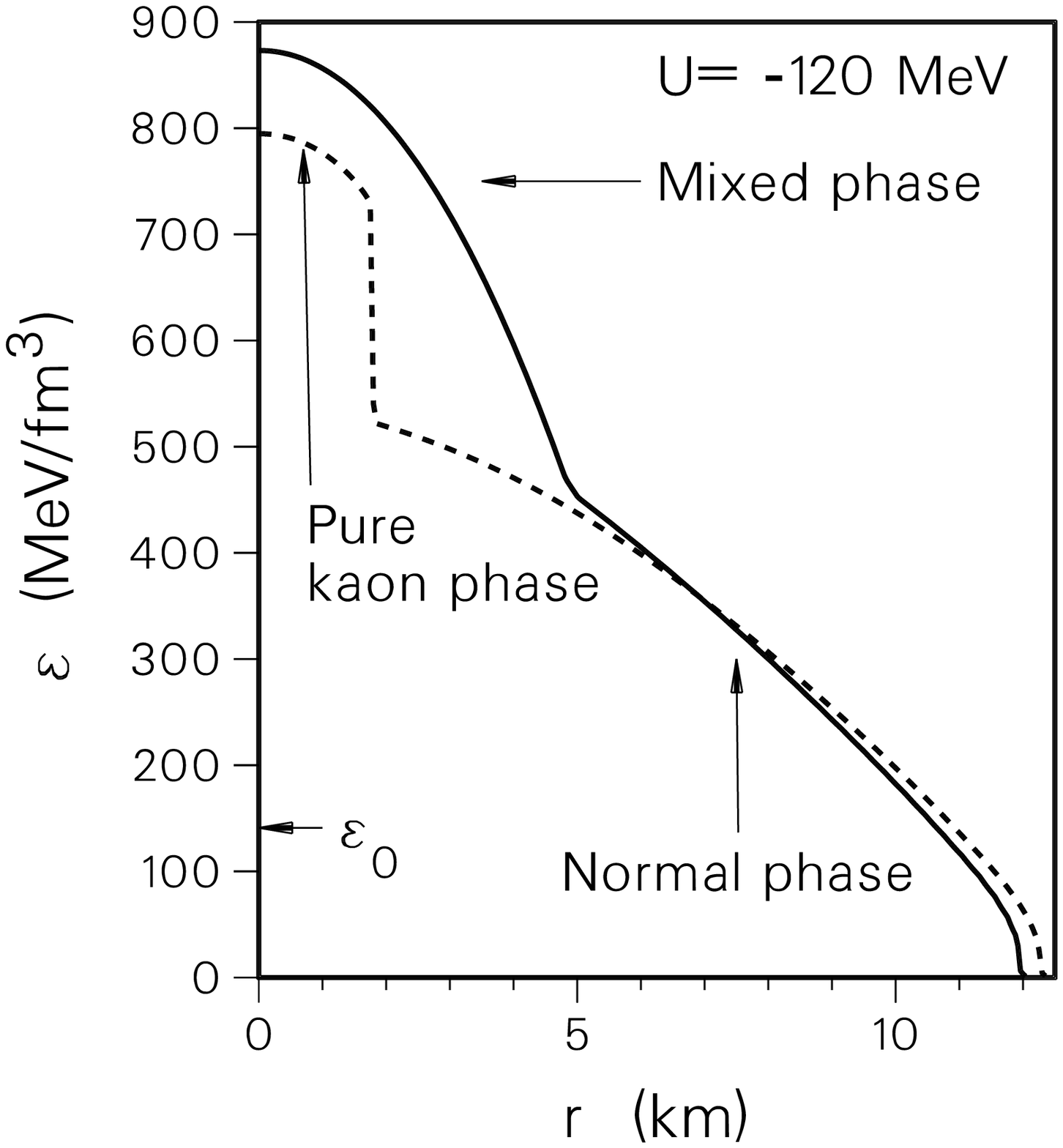,width=3.in}
\hspace{.1in}
\psfig{figure=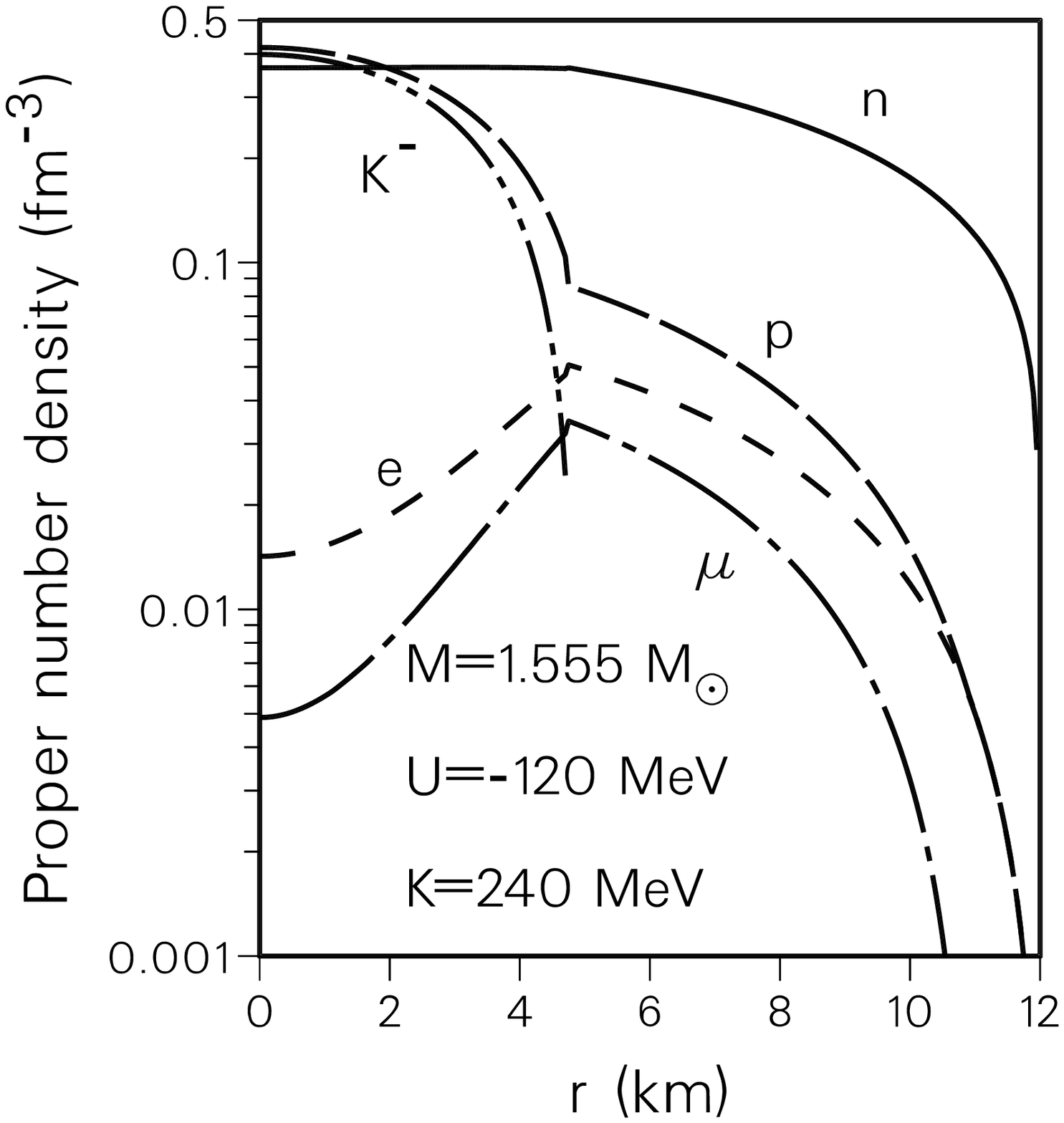,width=3in}
}}
\begin{flushright}
\parbox[t]{2.7in} { \caption { \label{prof_kaon_120} Mass-energy
distribution according to whether the mixed phase is treated
by the Maxwell construction (dashed line),
 or so as to respect the continuity of
 both chemical potentials (solid line).
}} \ \hspace{.4in} \
\parbox[t]{2.7in} { \caption { \label{comp_kaon_120}The composition 
of the maximum mass neutron star with a mass of $M=1.555$
M$_\odot$. Note that while protons are the dominant species at the
center of the star, overall, they are a minority population.
}}
\end{flushright}
\end{center}
\end{figure}

We have already stressed how differently the 
computed equation of state and matter
properties are, depending on whether the Maxwell construction is used to
determine (incorrectly) the mixed phase of normal and condensed phase, or
whether Gibbs criteria for equilibrium are fully respected. We start our
discussion of the large scale
properties of stars by illustrating the difference in
the mass-energy distribution in a star depending on which method is used. 
Fig.\ \ref{prof_kaon_120} shows the distribution in the two cases.
For the Maxwell
construction the energy density is discontinuous
at the particular radius at
which the pressure has the constant value of the  Maxwell construction.
The discontinuity is
analogous to the separation of the phases in a gravitational field that is
characteristic of a substance having a single component (like the steam above
water in H$_2$O). As we discussed earlier, neutron star matter in beta
equilibrium does not behave like that: it has two independent components
and all properties are continuous from one phase to another. The
distribution of mass-energy for such a star is the continuous curve with a
discontinuity in {\sl slope} 
but not {\sl value}, 
at the boundary between mixed phase. The central core of mixed phase
is surrounded by normal dense
nuclear matter. For the particular value of $U(\rho_0) = -120 $
MeV, the mixed phase extends to the center of the star and the pure
condensed phase does not appear.

Depending on the kaon potential $U_K(\rho_0)$, the pure kaon condensed phase
may not appear in the star, even for the star at the mass limit. Such is the
case in the above illustration.  However, for a potential $U_K(\rho_0)=-140$
MeV, the pure kaon condensed phase would form the core of stars with a mass
above about $1.1 M_\odot$, and for the limiting mass star, the condensed phase
would extend to about 4.5 km.

The distribution of particles in the limiting mass star is dominated by the
neutron in the normal phase outside 3 km as can be seen
from  Fig.\ \ref{comp_kaon_120}. The K$^{-}$ and proton are the
dominant species in the mixed phase core. Lepton populations fall rapidly, as
expected, as the K$^{-}$ becomes dominant.  However, overall, the
proton population is far less than the neutron, and there appears little
justification in referring to a star with a kaon condensate as a
nucleon star.

\begin{figure}[tbh]
\vspace{-.5in}
\begin{center}
\leavevmode
\centerline{ \hbox{
\psfig{figure=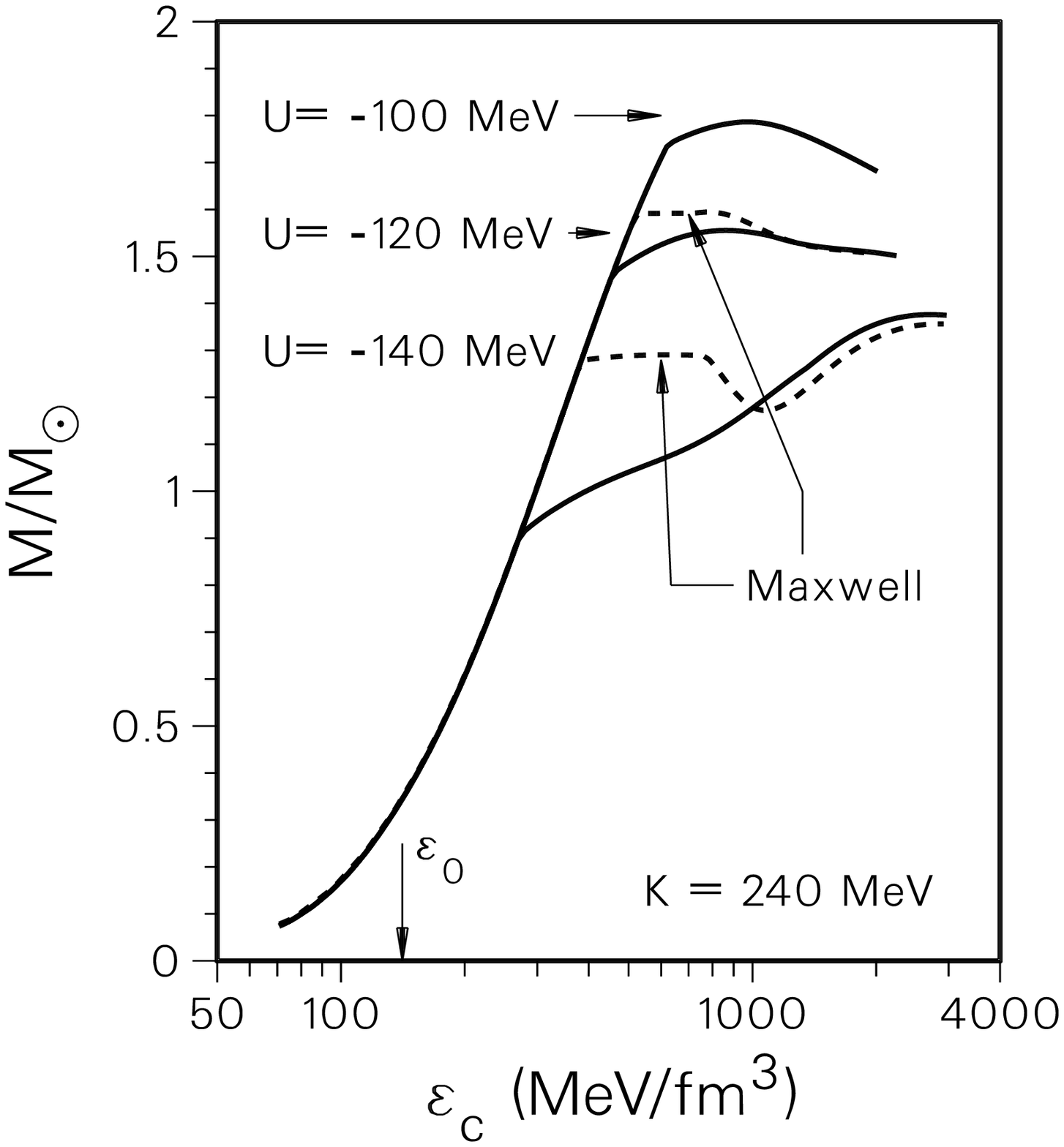,width=3.in}
\hspace{.1in}
\psfig{figure=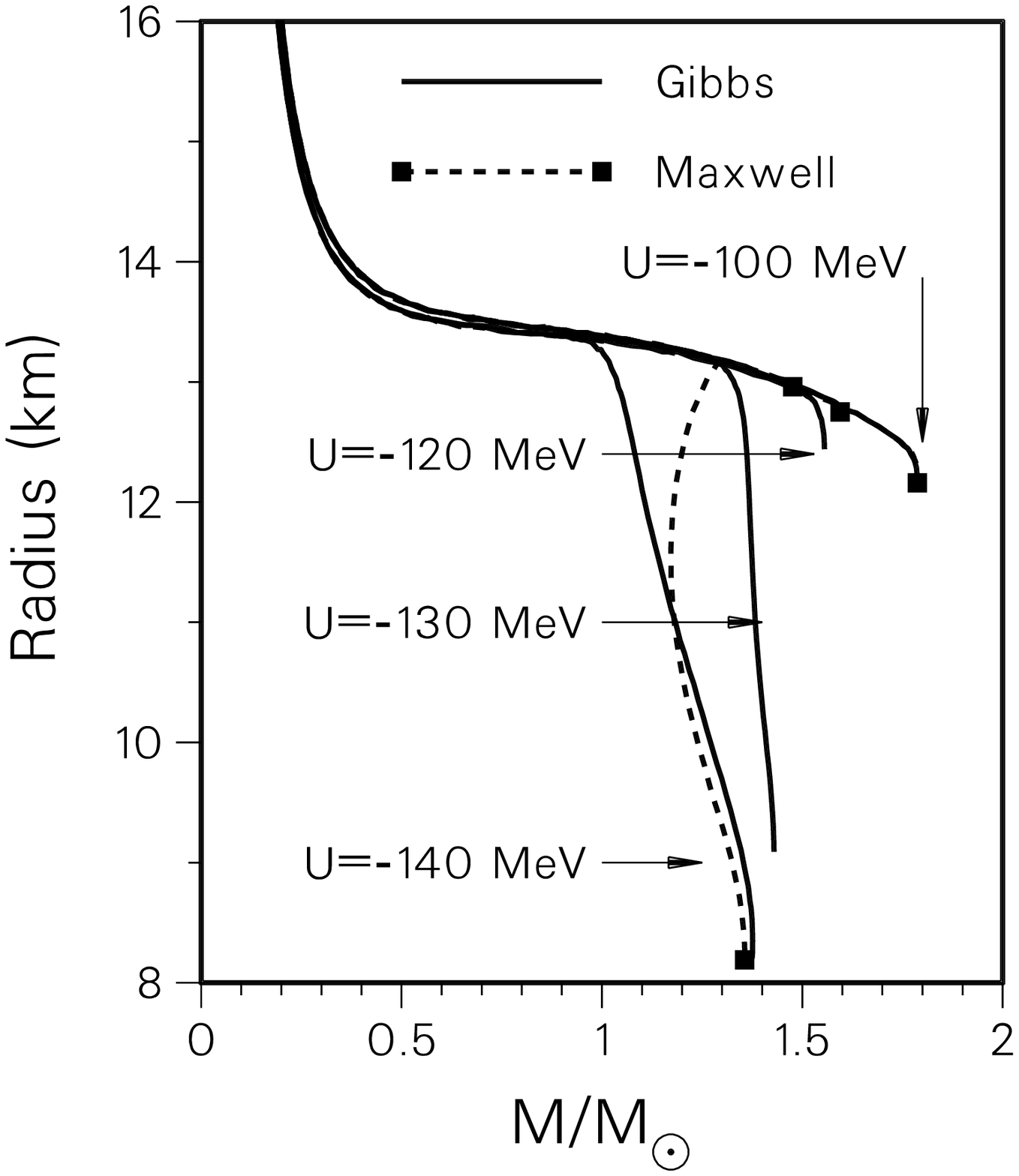,width=3in}
}}
\begin{flushright}
\parbox[t]{2.7in} { \caption {  \label{mas_kaon} The mass sequence for kaon
condensed neutron stars treated by the Maxwell construction (dashed line),
or so as to respect the continuity of
both chemical potentials (solid line).
}} \ \hspace{.4in} \
\parbox[t]{2.7in} { \caption {  \label{rm_kaon} The mass-radius 
relation for kaon
condensed neutron stars using a Maxwell construction (dashed line) or Gibbs
condition (solid line).
}}
\end{flushright}
\end{center}
\end{figure}

Stellar sequences for several choices of the kaon potential $U_K(\rho_0)$ are
shown in Fig.\ \ref{mas_kaon}. Naturally the limiting mass decreases with
increasing potential (for which the condensate density threshold is
lower). Potentials with values only a little below $U_K(\rho_0)=-120$ MeV would
not be compatible with the mass of the Hulse-Taylor pulsar, for the underlying
theory of matter used here. 
There is a mechanical instability  for the Maxwell case that
is initiated by the 
central densities for which the pressure remains
constant. In this case the {\sl necessary} condition for stability,
$dM/d\epsilon_c>0$ is not satisfied. (The section of the dashed curve
in Fig.\ \ref{rm_kaon} for which $R$ is an increasing function of $M$
is the corresponding unstable region.)
Such an unstable region is
absent when the phase transition is treated using Gibbs' conditions.

The mass-radius relation for several sequences is shown in Fig.\
\ref{rm_kaon}. Comparison is made with the case corresponding to the Maxwell
construction for the phase transition. It is clear that the radius especially
and the limiting mass are  sharp functions of $U_K(\rho_0)$. 
For the preferred value of $U_K(\rho_0)=-120$ MeV, radii are similar to neutron
stars without the condensate. There appears to be a sharp break in behavior
of $M$ vs. $R$ for $U_K(\rho_0)<-120$ MeV. However the behavior is actually
continuous but depends sensitively on $U_K(\rho_0)$: a pure quark core
develops with decreasing values of the optical potential below $\sim
-120 $ MeV and this causes the change of the radius from $R\approx 12.5$ km to
$R\approx 8$ km for $U_K(\rho_0)=-140$ MeV.

\subsection{Geometrical Structure in the Mixed Phase}
\label{sec:structure}

Neutron star matter in the normal phase is necessarily highly isospin
asymmetric since charge neutrality is imposed by the weakness of the
gravitational field compared to the Coulomb force.  
However, since kaons are bosons, they can all occupy the
zero momentum state. Consequently, when the two phases, normal and 
kaon condensed are in phase equilibrium, the normal phase can come
closer to isospin
symmetry as can be seen in Fig.\ \ref{fig:populmix120}.
This is achieved by charge exchange as driven by the isospin restoring
force arrising in part from the Fermi energies and in part from the
coupling of the $\rho$ meson to the nucleon isospin.
Naturally, the possibility of
achieving symmetry varies as the proportion of the kaon phase. Regions of
normal matter will be positively charged while regions of the kaon condensed
phase will be negatively charged.  Charge neutrality is globally achieved in
this way, but not locally.
As was discussed in Ref. \cite{glen91:a},
regions of like charge will tend to be broken up into small
regions while the surface interface energy will resist. The competition
is resolved by formation of a Coulomb lattice much as nuclei embeded in
an electron gas. The difference here is that it is two phases of nuclear
matter that are involved. The rarer phase will occupy lattice
sites embedded in the dominant phase. As the proportion of phases changes,
the total energy consisting of  volume, surface and Coulomb energies 
will be minimized by a sequence of geometrical
forms at the lattice sites, which we idealize as drops, rods and
slabs, just as for nuclear matter embedded in a background of 
free electrons and neutrons
\cite{ravenhall83:aa}.

\begin{figure}[tbh]
\vspace{-.5in}
\begin{center}
\leavevmode
\centerline{ \hbox{
\psfig{figure=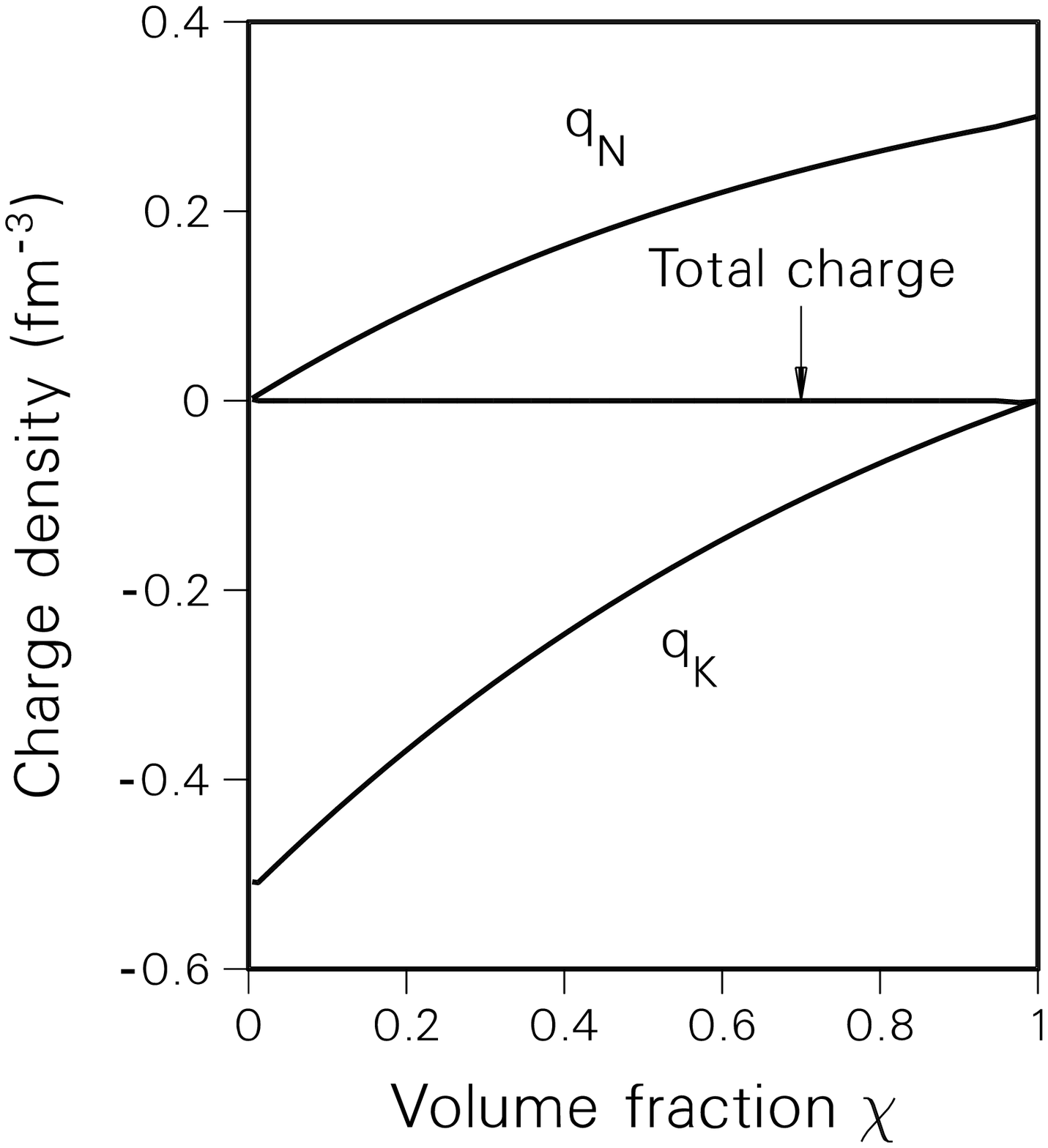,width=3.in}
\hspace{.1in}
\psfig{figure=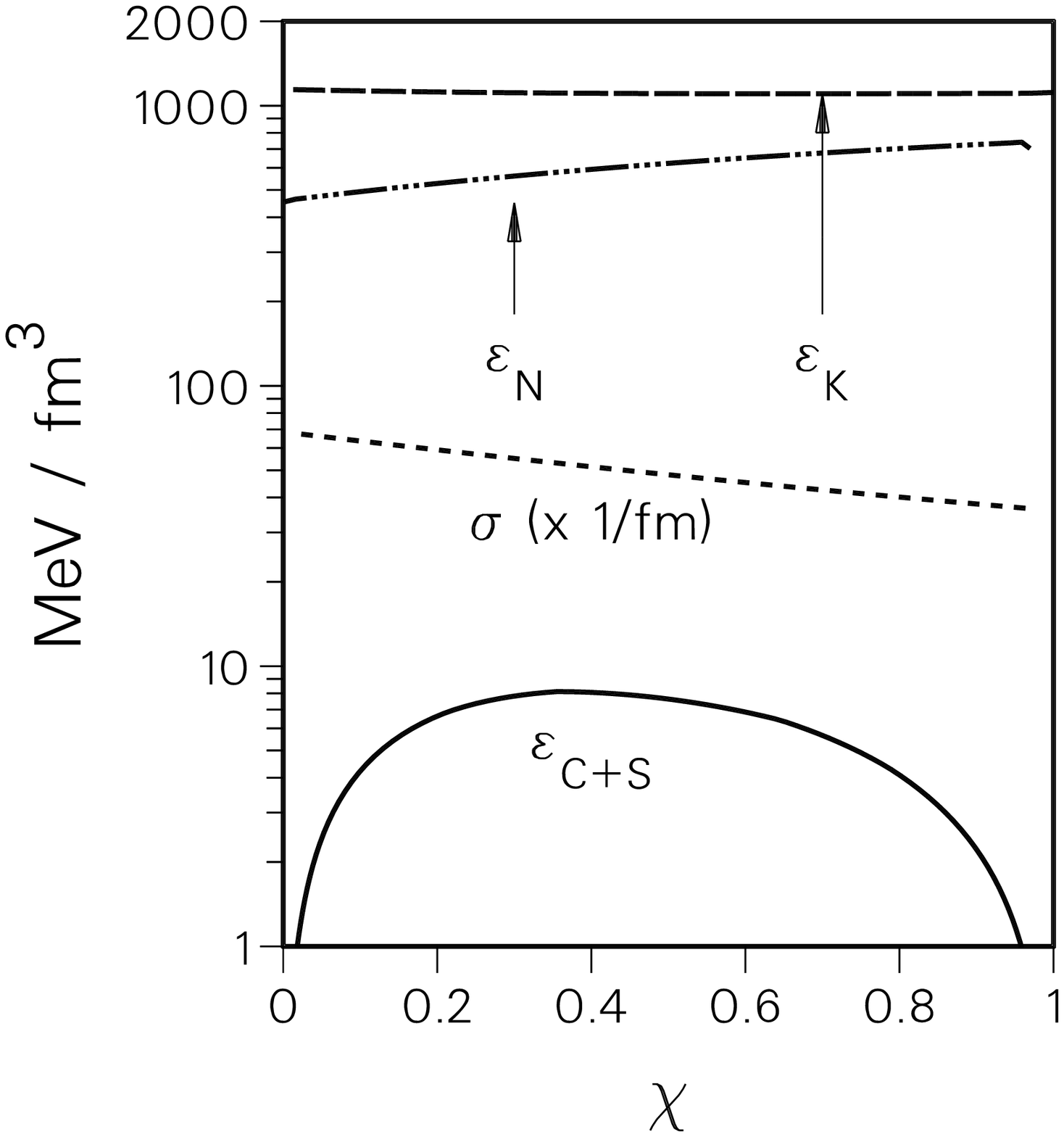,width=3in}
}}
\begin{flushright}
\parbox[t]{2.7in} { \caption { \label{chiq_kaon_120} Charge densities
in the normal and kaon condensed phase as a function of the volume fraction
of the latter.
}} \ \hspace{.4in} \
\parbox[t]{2.7in} { \caption { \label{prop_kaon_120}  Bulk
energy densities
of normal and kaon phases as a function of the volume fraction of the
kaon phase, the surface tension $\sigma$ which is assumed to be proportional
to their difference, and the sum of Coulomb and surface energy density.
}}
\end{flushright}
\end{center}
\end{figure}

\begin{figure}[tbh]
\vspace{-.5in}
\begin{center}
\leavevmode
\centerline{ \hbox{
\psfig{figure=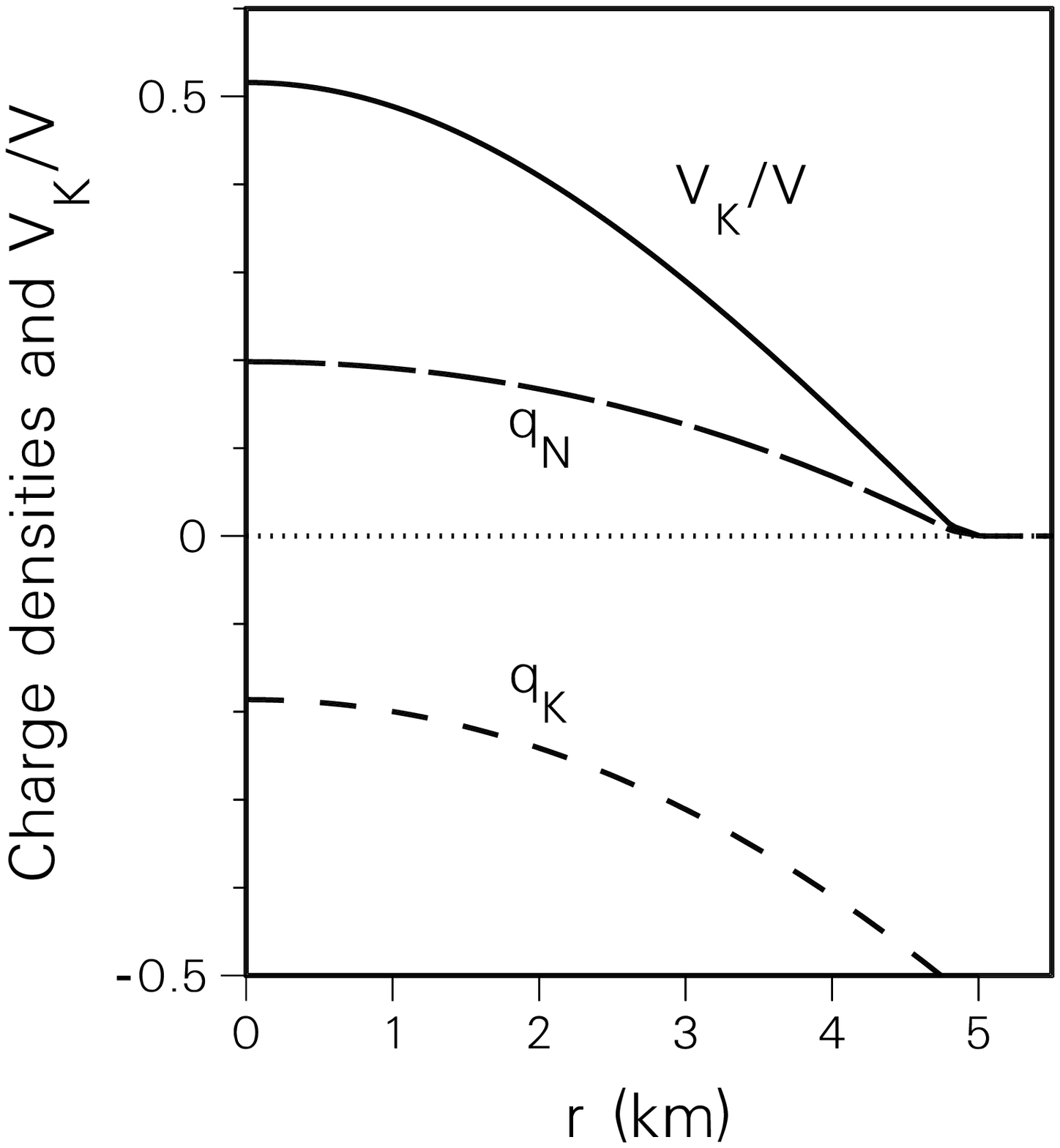,width=3.in}
\hspace{.1in}
\psfig{figure=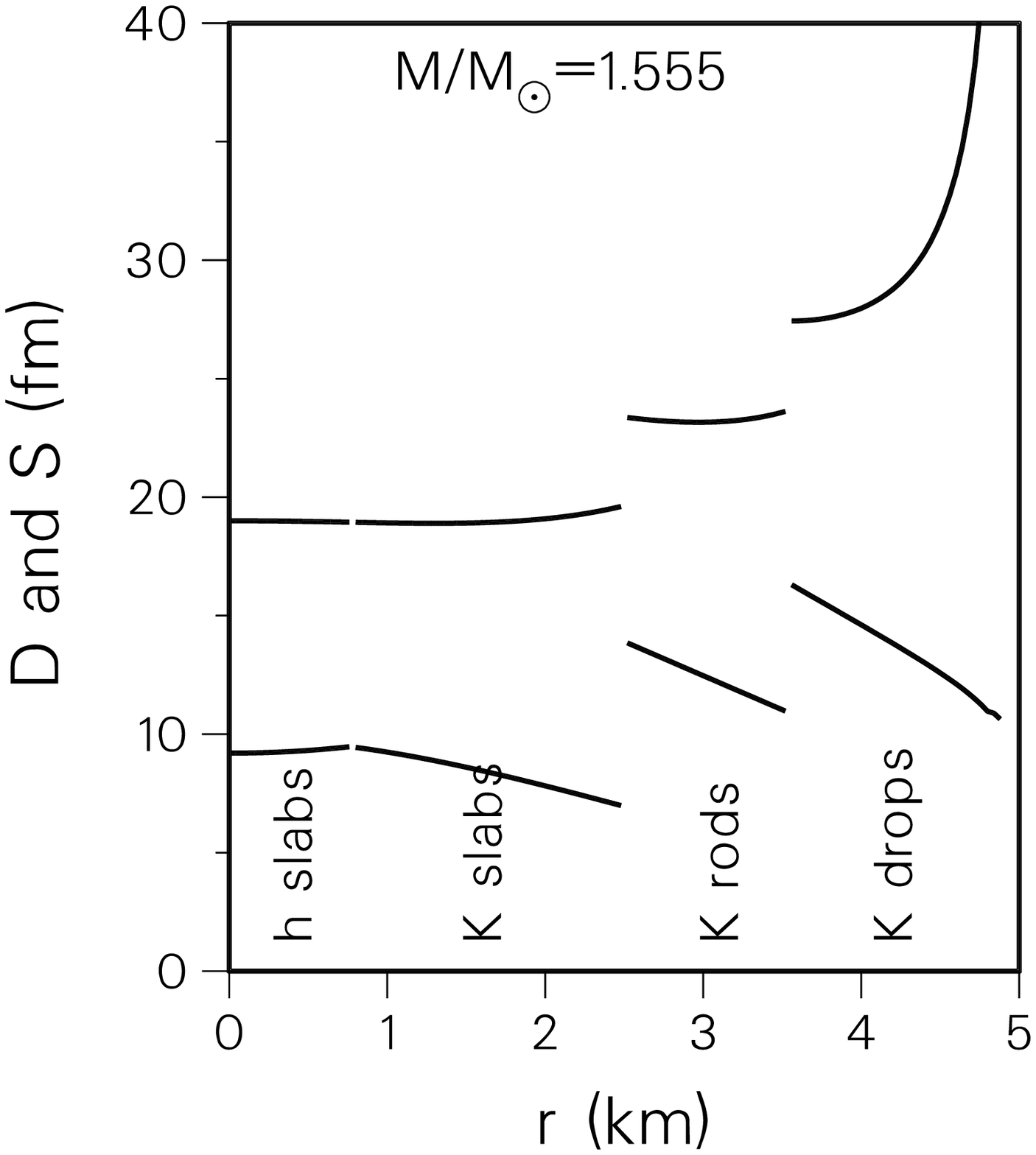,width=3in}
}}
\begin{flushright}
\parbox[t]{2.7in} { \caption { \label{chi_kaon_120}Charge densities in
the normal and kaon condensed fractions of the mixed phase in the
limiting mass star of the case $U(\rho_0)=-120$ MeV. Volume fraction
of kaon phase is also plotted. 
}} \ \hspace{.4in} \
\parbox[t]{2.7in} { \caption { \label{crys_kaon_120} Diameter D of objects
(drops, rods, slabs) of the rarer phase immersed in the dominant
phase, located at lattice sites
spaced  S apart.
}}
\end{flushright}
\end{center}
\end{figure}

Relevant details of the structure calculation can be found in Ref.
\cite{glen95:c,glen95:e}. In the present situation, the physical quantities 
that determine the  geometrical structure are shown in 
Figs.\ \ref{chiq_kaon_120} and \ref{prop_kaon_120}. 
Of course a calculation of the geometric structure, which results from a
competition between Coulomb and surface energies, requires a knowledge
of the surface tension $\sigma$
at the interface between the phases.
This is not known although
a calculation is in progress
\cite{christiansen99:a}.
What we do know is that: (1) The sizes, spacings and
the sum of Coulomb and surface energies scale as $\sigma^{1/3}$.
(2) To first approximation, the locations of the transition from
one geometric phase to another does not depend on $\sigma$. The reason for 
this is that the  sum of Coulomb and surface energy densities is small
compared to the bulk energy 
density (cf. Fig.\  \ref{prop_kaon_120}).
(3) The threshold density of the mixed phase and the density at which
it ends is not disturbed by our uncertainty in $\sigma$ because
the sum of Coulomb and surface energies vanishes at the end points
(Fig.\ \ref{prop_kaon_120}, see  eq. (2) in Ref.
\cite{glen95:c}).
(4) The structured phase lies lower in energy than the unstructured
 (See near end of introduction of Ref. \cite{glen97:c}.) 
For the above reasons the dimensions shown in Fig.\ \ref{crys_kaon_120}
provide a guide but the locations of phases should be quite
accurate.

\begin{figure}[tbh]
\vspace{-.5in}
\begin{center}
\leavevmode
\centerline{ \hbox{
\psfig{figure=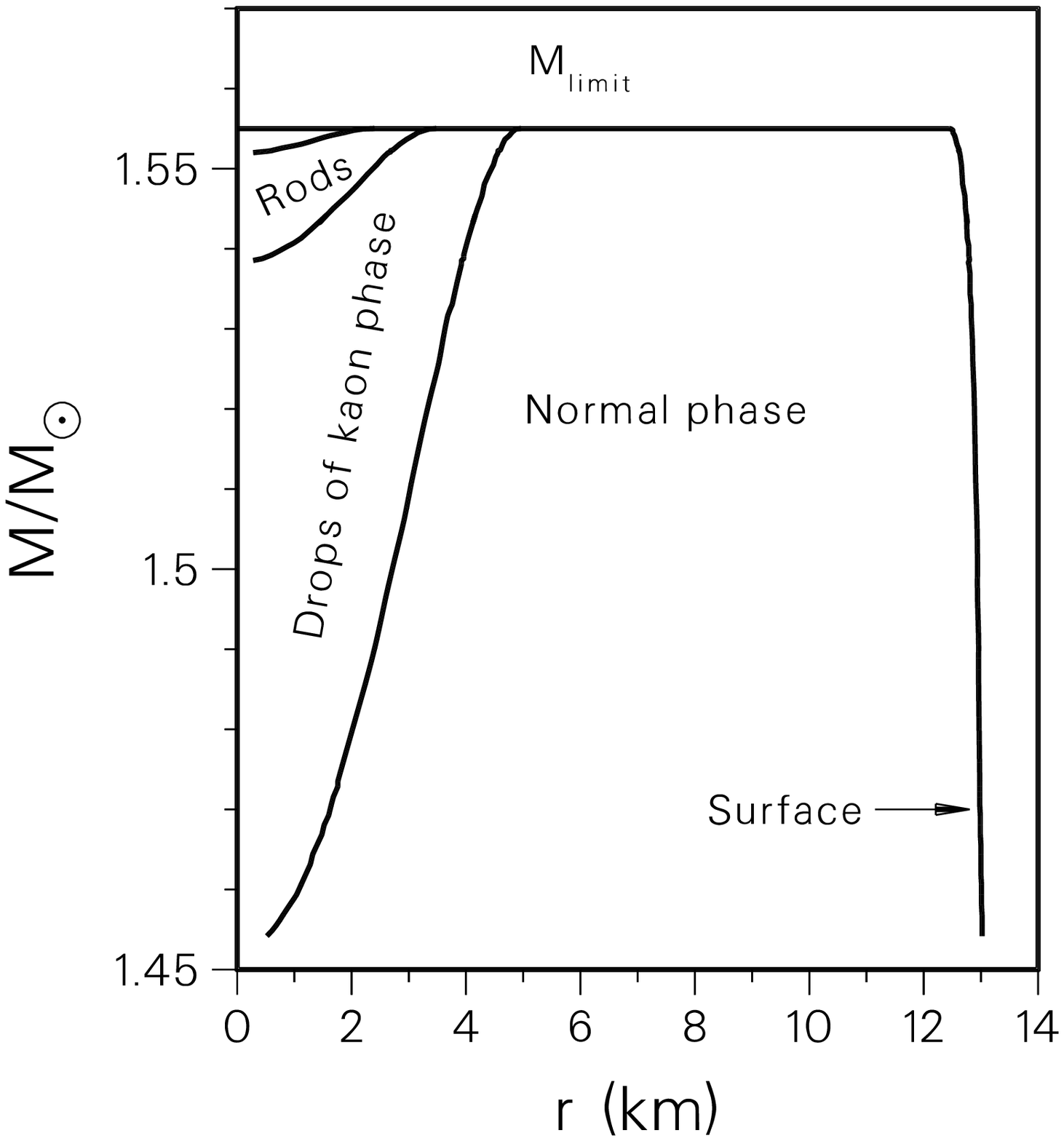,width=3.in}
\hspace{.1in}
\psfig{figure=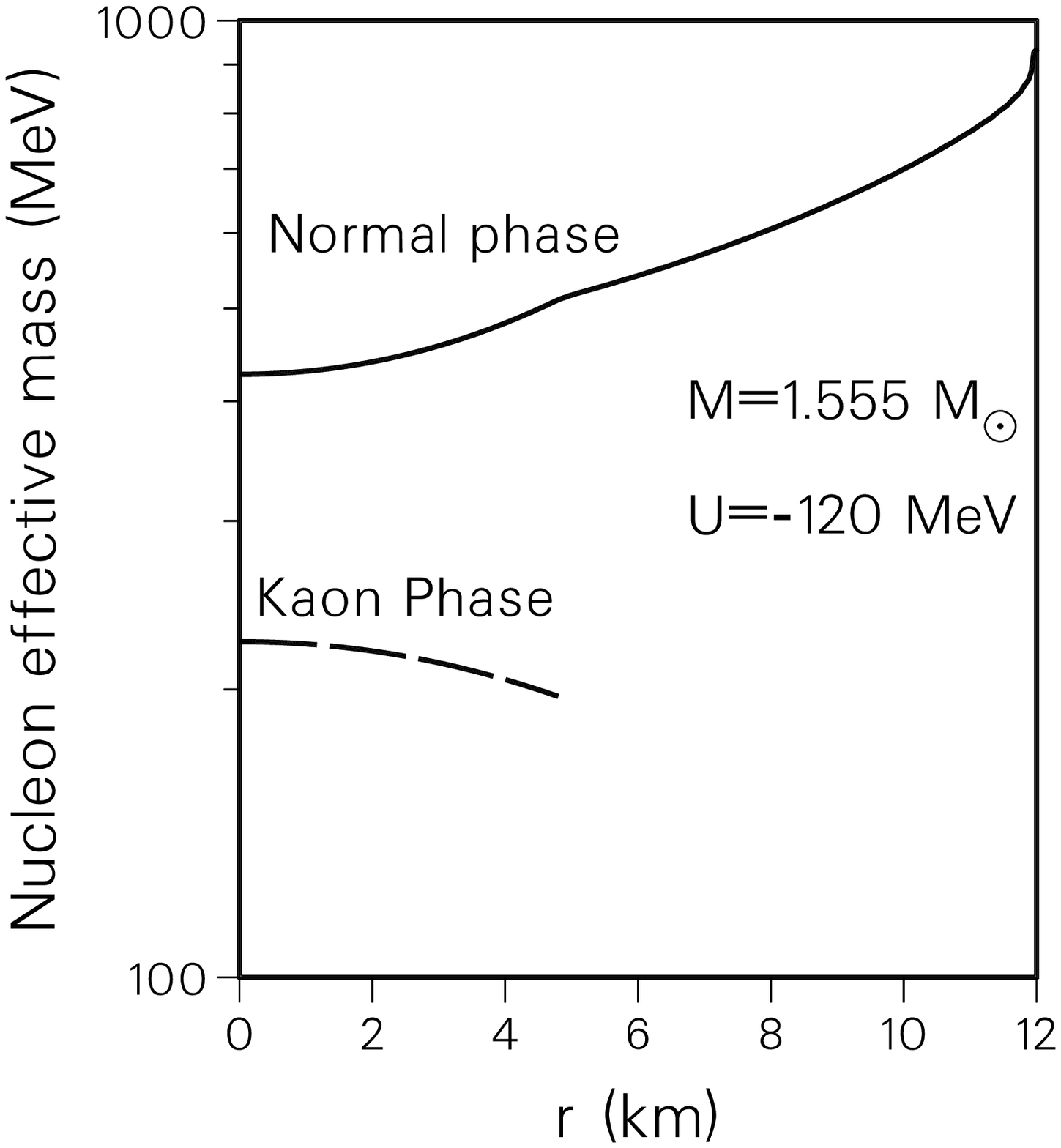,width=3in}
}}
\begin{flushright}
\parbox[t]{2.7in} { \caption { \label{contour_kaon} Radial
boundaries between phases
are shown for a range of stellar masses.
}} \ \hspace{.4in} \
\parbox[t]{2.7in} { \caption { \label{effmass_kaon_120} Nucleon
effective mass in the normal and kaon condensed phase as a function
of radial location in the limiting mass star.
}}
\end{flushright}
\end{center}
\end{figure}

The charge densities carried by the two phases and the volume
fraction of the kaon phase is shown in Fig.\ \ref{chi_kaon_120} as a function
of radial coordinate in the star. Outside of 5 km, matter is in the pure
nuclear phase and it is chargeless, the proton population being balanced by
electrons.
In the idealized geometry of shapes, the kaon phase will
first form at the threshold density of condensation as spheres
spaced far apart. As the fraction of kaon phase increases, the spacing will
decrease and eventually the spheres will merge to form rods and then slabs. As
the volume fraction of the kaon phase comes
to dominate, slabs of normal phase will be present in a background of kaon
phase, and the role of the two phases is interchanged.
The diameter and spacing
of the geometrical forms of the crystal lattice is shown in
Fig.\ \ref{crys_kaon_120}
for the limiting mass star. The location of the boundaries of the
various phases can be seen in Fig.\ \ref{contour_kaon}
for stars of various mass.

These are rather remarkable properties of the mixed phase, which in the model
star, occupies the inner 5 km. It is filled with geometrical forms of varying
shapes and spacings, according to depth in the star. The charge density within
the geometrical objects and the background phase is opposite in sign and
varying in magnitude with depth. Finally the effective mass of the nucleons is
radically different in the two phases as can be seen in
Fig.\ \ref{effmass_kaon_120}. All of these features must have their effect on
transport properties and possibly on Glitch phenomena. Glitches are thought to
correspond to changes induced in the moment of inertia of the star as a massive
number of superfluid vortex lines undergo  shifts 
in the location  of the sites in the solid regions to which they are
pinned \cite{Alpar96}. 
The relocation occurs unpredictably as the instantaneous
location of the vortices
carrying the angular momentum come out of equilibrium with  
decreasing spin of the star and create stresses that are relieved by the
massive unpinnings. The thin crust is a location at which the
vortex lines can be pinned. But in the present model, the vortex lines
do not thread through the entire star, pinned at each end on the crust,
but are pinned at one end on the interior crystalline mixed phase.
The extent of this region varies sensitively as the mass of the star,
perhaps accounting for the wide variety in glitch phenomena observed in
different pulsars.


\section{Summary}
\label{sec:summary}

We have discussed the properties of kaon condensation in neutron star matter
when it is of first order. As is general for any phase transition in a 
substance having more than a single conserved charge, the mixed phase does 
occupy a finite extent in the star, and in that region is quite rich in 
phenomena. First, a  Coulomb lattice of rare phase immersed in the
dominant one will will form, having various geometries at the
lattice sites, according to the pressure.
This feature  is common to nuclear systems having a mixed phase
(independent of the phase transition) so long as the
temperature is low on the nuclear scale.
Second, nucleons have different mass
depending on whether
they are in the objects at the lattice sites, or in the background
medium. Third, the objects at the lattice sites have opposite charge
compared to the background. Thus the mixed phase region, which we calculate
to occupy a region of a few kilometers in extent, is highly heterogeneous.
We believe this will be an important factor in determining the transport
properties of this region. Moreover, the solid region, if present,
is likely to play a 
role in the pulsar glitch phenomenon and its extent in the core, being very
sensitive to stellar mass, may account for the variety of
glitch phenomena observed in different pulsars.

\section*{Acknowledgments}

J. S.-B. acknowledges support by the Alexander von Humboldt-Stiftung.
This work was supported by the Director, Office of Energy Research, Office of
High Energy and Nuclear Physics, Division of Nuclear Physics, of the U.S.
Department of Energy under Contract DE-AC03-76SF00098.


\end{document}